\documentclass[11pt]{article}
\usepackage{amsmath,amsfonts}
\usepackage{mathtools}

\usepackage{graphicx}
\usepackage{epstopdf}
\usepackage{amsmath}
\usepackage{amssymb}
\usepackage{amsfonts}
\usepackage{amsthm}
\usepackage[usenames]{color}
\usepackage{array}
\usepackage{verbatim}
\usepackage{soul}
\usepackage{braket} 
\usepackage{xcolor}
\usepackage[
      colorlinks=true,
      linkcolor=blue,
      urlcolor=blue,
      filecolor=blue,
      citecolor=red,
      pdfstartview=FitV,
      pdftitle={},
      pdfauthor={},
      pdfsubject={},
      pdfkeywords={},
      pdfpagemode=None,
      bookmarksopen=true
]{hyperref}

\usepackage{epsfig}

\usepackage{hyperref}
\newcommand{\vac}{\ket{\varnothing}}

\newcommand \nn{\nonumber\\}

\def\be{\begin{equation}}
\def\ee{\end{equation}}
\def\bea{\begin{eqnarray}}
\def\eea{\end{eqnarray}}

\def\be{\begin{equation}}
\def\ee{\end{equation}}
\def\bea{\begin{eqnarray}}
\def\eea{\end{eqnarray}}

\newcommand{\rom}[1]{\mathrm{#1}}

\def\cA{\mathcal{A}}

\def\cV{\mathcal{V}}

\def\la{\lambda}

\numberwithin{equation}{section}

\begin{document}

\begin{centering}

\textbf{\LARGE{Quasinormal modes of supersymmetric microstate \\  \vspace{0.2cm} geometries from the D1-D5 CFT}}

 \vspace{0.8cm}
 Bidisha Chakrabarty$^1$, Debodirna Ghosh$^2$, Amitabh Virmani$^2$
  \vspace{0.5cm}

 \vspace{0.5cm}

\begin{minipage}{.9\textwidth}\small  \begin{center}
$^1$International Centre for Theoretical Sciences (ICTS), \\
Survey No. 151, Shivakote, 
Hesaraghatta Hobli, 
Bengaluru - 560 089, India \\

 \vspace{0.5cm}

$^2$Chennai Mathematical Institute, \\ H1 SIPCOT IT Park, Kelambakkam, Tamil Nadu 603103, India \\
  \vspace{0.5cm}
{\tt bidisha.chakrabarty@icts.res.in, debodirna@cmi.ac.in,  avirmani@cmi.ac.in}
\\ $ \, $ \\
\end{center}
\end{minipage}

\end{centering}

\begin{abstract}
We revisit the study of the probe scalar quasinormal modes of a class of three-charge super- symmetric microstate geometries. We compute the real and imaginary parts of the quasinormal modes and show that in the parameter range when the geometries have large AdS region, the spectrum is precisely reproduced from a D1-D5 orbifold CFT analysis. The spectrum includes the slow decaying modes pointed out by Eperon, Reall, and Santos. We analyse in detail the nature of the quasinormal modes by studying the scalar wavefunction. We show that these modes correspond to slow leakage of excitation from AdS throat to infinity. 
\end{abstract}

\newpage

\tableofcontents
\newpage

\section{Introduction}

The fuzzball proposal \cite{Lunin:2001jy, Mathur:2005zp, Bena:2007kg, Skenderis:2008qn, Bena:2013dka} is a paradigm for the black hole interior. In the conventional picture of a black hole, the region around the horizon is in the vacuum state. In the fuzzball paradigm, string theory effects modify the interior of the black hole all the way upto the horizon. The proposal posits that the radiation leaving from the fuzzball can carry information just like radiation from a piece of burning coal. In recent years, these ideas are widely explored, though still there are many issues that remain to be understood, particularly with regard to how the fuzzball proposal reproduces detailed classical and quantum properties of  black holes. 

In the context of supersymmetric D1-D5-P black hole, the fuzzball program has met with the most success. Large classes of supersymmetric microstate geometries have been constructed. These supersymmetric microstate geometries provide a geometrical description of certain quantum microstates of the black hole. In several cases their identification in the dual CFT as states is also well understood.  The focus of attention in the present work is the class of geometries constructed by Giusto, Mathur, and Saxena (GMS)~\cite{gms1,gms2, Giusto:2004kj}. These geometries carry three charges as well as two angular momenta. These geometries can be thought of as special limits of the Cveti\v{c}-Youm metrics.

The GMS microstate geometries have been studied in a variety of contexts over the years. These geometries admit a Killing tensor in addition to four Killing vectors in 6d. As a result geodesic equations can be reduced to quadratures and the scalar wave equation can be separated. Smooth  (linear as well as non-linear) hair modes on the dilaton-free class of GMS geometries have been constructed in the last decade or so~\cite{Mathur:2003hj, Mathur:2011gz, Lunin:2012gp, Giusto:2013rxa}. More recently, via a tour de force calculation (non-linear) smooth hair modes on general GMS geometries supporting non-trivial dilaton profiles have also been constructed~\cite{Chakrabarti:2019lfu}. The dual CFT description of these hair modes is also understood. Non-supersymmetric generalisation of the GMS geometries, known as the JMaRT solutions~\cite{Jejjala:2005yu}, are also well explored~\cite{Cardoso:2005gj, Chowdhury:2007jx, Katsimpouri:2014ara, Chakrabarty:2015foa}.

In an intriguing paper~\cite{Eperon:2016cdd}, Eperon, Reall, and Santos (ERS) studied linear waves in the GMS microstate geometries, and pointed out that there are qualitative differences between the decay of linear waves in supersymmetric microstate geometries and in supersymmetric black holes. In supersymmetric black holes linear waves for scalar perturbations decay as an inverse power of time at late times. The higher angular frequency modes decay faster. In contrast,  in the GMS microstate geometries there is a stable trapping of geodesics.  The stable trapping of geodesics strongly suggests that the decay of linear waves is slower than any inverse power of time. To further support this claim they pointed out that the GMS geometries admit very slow decaying quasinormal modes. The slower decaying modes are with higher angular frequencies. They argued that such a  slow decay leads to a non-linear instability of mircostate geometries.\footnote{Geodesics in the GMS  microstate geometries are further explored in~\cite{Eperon:2017bwq} and the decay properties of waves are further explored in~\cite{Keir:2016azt, Keir:2018hnv}.}

If we take the point of view that the fuzzball paradigm of black holes should reproduce classical properties of black holes in detail, then these results pose puzzles. Although subsequent authors have made suggestions for the end point of the ERS instability~\cite{Marolf:2016nwu, Bena:2018mpb}, several puzzles raised by ERS remain unaddressed. Specifically, how to reconcile with properties of black holes the fact that in the microstate geometries the slower decaying modes are of higher angular frequencies?  How to interpret the slow decaying quasinormal modes on the microstate geometries? In this paper, we present our modest attempts to address some of these issues.  The rest of the paper is organised as follows.

We  revisit the study of probe scalar quasinormal modes in the GMS microstate geometries in section \ref{sec:QNMs}. Since these geometries do not have a horizon,  quasinormal modes refer to definite frequency modes that satisfy regularity in the interior and are purely outgoing at infinity. We compute the real and imaginary parts of the quasinormal modes in the parameter regime when the geometries have large AdS region in the interior. In section \ref{sec:CFT} we reproduce both the real and the imaginary parts of the quasinormal mode spectrum from a D1-D5 orbifold CFT analysis. The CFT picture suggests a different scenario on the gravity side than the one suggested by ERS. They suggested that the slow decaying modes cause non-linear instability in the sense that a generic perturbation would collapse to a tiny black hole in the fuzzball geometry. Our results suggest a picture that the geometry evolves to another fuzzball upon scalar absorption. We conclude this because on the CFT side the scalar \emph{emission}  spectrum is reproduced via a transition from certain excited states to the GMS states. Unfortunately, the bulk duals for the excited CFT states are not known beyond the linear approximation. However, given our experience with fuzzball geometries and their CFT duals it is expected that the excited CFT states correspond to other fuzzballs, possibly with stringy microstructure.  This picture is in line with the suggestions made in \cite{Marolf:2016nwu, Bena:2018mpb}.

In section \ref{sec:wave_func} 
we analyse the quasinormal mode wavefunctions and certain features of the emitted scalar radiation. We make precise the physical picture that the decay process corresponds to the leakage of excitation from AdS throat to infinity. We conclude in section \ref{sec:conclusions}.  A collection of useful formulae for ease of reproducing calculations in the main text is presented in appendix \ref{app:compendium}.

The key reason for the appearance of the slow decaying modes in the GMS geometries is the ``reflecting'' boundary conditions. This boundary condition ensures that the scalar field is smooth at the location where the throat pinches to zero size. It is clearly the most natural boundary conditions for a probe scalar field on the classical microstate geometry. Given these boundary conditions the results of ERS are robust.  However, from the point of the view of the fuzzball proposal the smoothness of the scalar field or the ``reflecting'' boundary conditions is not the correct one. This point is closely related to the discussion presented in~\cite{Mathur:2012jk, Guo:2017jmi}. They argue that the dynamics of the scalar field on the microstate geometry is governed by the evolution of new excitations that the scalar field generates. These excitations alter the microstate geometry. The reflecting boundary condition overlooks this physics. Unfortunately, it is not clear how to account for such an effect in a probe calculation.

\section{Quasinormal modes of supersymmetric microstate geometries}
\label{sec:QNMs}
We begin with  a study of the quasinormal modes of the three-charge supersymmetric microstate geometries of \cite{gms1, gms2, Giusto:2004kj}. The term ``quasinormal modes'' simply refers to  modes with definite frequency $\omega$. We begin with a review of the scalar wave equation on the GMS geometries \cite{Giusto:2012yz, Lunin:2001dt, Eperon:2016cdd} and obtain the quasinormal mode spectrum via a matched asymptotic expansion analysis. Our investigation is inspired by the study of Eperon, Reall, and Santos \cite{Eperon:2016cdd}, who have also studied quasinormal mode in a matched asymptotic expansion. The main difference from their work is that our matched asymptotic expansion analysis is done in a ``near decoupling limit'', whereas their matched asymptotic expansion is done in the ``eikonal limit'' with the angular momentum parameter $l$ being large. In section \ref{ERS_limit} we present a detailed comparison of our expressions with theirs.
 
\subsection{Scalar wave equation}
The ten-dimensional string frame metric of the three-charge geometries of references \cite{gms1, gms2, Giusto:2004kj} takes the form,
\bea
\label{extremalmetric}
ds^2 &=& ds^2 _6 +\sqrt{H_1\over H_5}\sum_{i=1}^4 dx_i^2  \label{metric10d}
\eea
where
\bea
ds_6^2 & = & -\frac{1}{h} (dt^2-dy^2) + \frac{Q_{p}}{h
f}\left(dt-dy\right)^{2}+ h f \left( \frac{dr^2}{r^2 +
(\gamma_1+\gamma_2)^2\eta} + d\theta^2
\right)\nonumber \\
         &+& h \Bigl( r^2 + \gamma_1\,(\gamma_1+\gamma_2)\,\eta -
\frac{Q_1 Q_5\,(\gamma_1^2-\gamma_2^2)\,\eta\,\cos^2\theta}{h^2 f^2}
\Bigr)
\cos^2\theta d\psi^2  \nonumber \\
&+& h\Bigl( r^2 + \gamma_2\,(\gamma_1+\gamma_2)\,\eta +
\frac{Q_1 Q_5\,(\gamma_1^2-\gamma_2^2)\,\eta\,\sin^2\theta}{h^{2} f^{2}
}
\Bigr) \sin^2\theta d\phi^2  \nonumber \\
&+& \frac{Q_p\,(\gamma_1+\gamma_2)^2\,\eta^2}{h f}
\left( \cos^2\theta d\psi + \sin^2\theta d\phi \right)^{2} \nonumber\\
&-& \frac{2 \sqrt{Q_{1}Q_{5}} }{hf}
\left(\gamma_1 \cos^2\theta d\psi + \gamma_2 \sin^2\theta d\phi\right)
(dt-dy)
\nonumber \\
&-& \frac{2 \sqrt{Q_1 Q_5}\,(\gamma_1+\gamma_2)\,\eta}{h f}
\left( \cos^2\theta d\psi + \sin^2\theta d\phi \right) dy.  \label{metric6d}
\eea
Explicit expressions for the 2-form Ramond-Ramond field and the dilaton can be found in the above references. These geometries are solutions to the type IIB supergravity compactified on T$^4$.  Upon dimensional reduction, the resulting six-dimensional geometries are asymptotically 5d Minkowski spacetime times a Kaluza-Klein circle of radius $R$. 
We will focus on the six-dimensional geometry. Various functions and parameters appearing in metric \eqref{metric10d}--\eqref{metric6d} are as follows,
\bea
&&\!\!\!\!\!\!\!\!\!\!\!\!\eta ~=~ {Q_1 Q_5\over Q_1 Q_5 + Q_1 Q_p + Q_5
Q_p}, \\
&&\!\!\!\!\!\!\!\!\!\!\!\!f ~=~ r^2+ (\gamma_1+\gamma_2)\,\eta\,
\bigl(\gamma_1\, \sin^2\theta + \gamma_2\,\cos^2\theta\bigr), \\
&&\!\!\!\!\!\!\!\!\!\!\!\!H_{1} ~=~ 1+
\frac{Q_{1}}{f}\,,\quad H_{5} ~=~ 1+ \frac{Q_{5}}{f}\,,\quad h ~=~
\sqrt{H_{1} H_{5}},
\label{deffh}
\\
&& \!\!\!\!\!\!\!\!\!\!\!\! Q_p=
-\gamma_1 \,\gamma_2.
\label{condition}
\eea

The $y\sim y + 2 \pi R$ circle plays a key role in our discussion. In writing the above metric we have in mind the D1-D5 system, that is,  we  consider $n_1$ D1 branes wrapped on this $S^1$ and $n_5$ D5 branes  wrapped on $S^1 \times T^4$. The parameters $Q_1, Q_5$ and $Q_p$ are the dimensionful charges, which can be written in terms of integer charges $n_1, n_5$ and $n_p$ as follows,
\be
Q_1 = \frac{g_s \alpha'{}^3}{V}n_1, \qquad Q_5 = g_s \alpha' n_5, \qquad Q_p = \frac{g_s^2 \alpha'{}^4}{V R^2}n_p,
\ee
where the volume of the $T^4$ at infinity is $(2\pi)^4 V$. 

It is convenient to  work with the following parametrisation,
\be
\gamma_1 =  -{\sqrt{Q_1 Q_5}\over R}\,n \gamma \,,\quad \gamma_2 =
{\sqrt{Q_1 Q_5}\over R}\,(n+1)\gamma,\quad Q_p  ={Q_1 Q_5\over R^2}\,
n(n+1)\gamma^2=a^2 n(n+1)\gamma^2
\label{gamma2}
\ee
where 
\bea
a=\frac{\sqrt{Q_1Q_5}}{R} , \label{parameter_a}\qquad
\gamma = \frac{1}{k}.
\eea
The parameter $k \in \mathbb{Z}$ is the orbifolding parameter. The orbifold structure of these geometries was analysed in detail in reference \cite{Giusto:2012yz}. 
From above relations it follows that the integer quantised momentum charge $n_p$ is, 
\be
n_p =  n (n+ 1) n_1 n_5 \gamma^2.
\ee

A minimally coupled scalar on the six-dimensional (Einstein frame) metric $ds_6^2$ has a natural interpretation in IIB supergravity as a graviton with legs in the T$^4$ directions; see e.g.,~\cite{David:2002wn}. 
The equation for the minimally coupled massless scalar,
\be
\Box\,\Phi \equiv  {1\over \sqrt{-g}}\,\partial_\mu\,
\Bigl( \sqrt{-g}\,g^{\mu\nu}\,\partial_\nu\,\Phi\Bigr)=0,
\ee
 is known to separate for scalar configurations of the form 
\be
\Phi(t,r,\theta,\phi,\psi,y)=\exp(-i \omega t+ i m_{\phi}\phi+i m_{\psi}\psi+i \la y)
H(r)\Theta(\theta). \label{scalar_ansatz}
\ee
Our conventions are same as \cite{Chowdhury:2007jx, Chakrabarty:2015foa}, except that $\lambda$ now has dimensions of inverse length. Positive $\omega_R$ corresponds to positive energy quanta. Upon substituting this ansatz, we get the angular equation to be 
\bea
&&\!\!\!\!\!\!\!\!\!\!\!\!\!\!\!\!\!\!{1\over \sin 2\theta}{d\over
d\,\theta}\left(\sin 2\theta {d\over d\,\theta}
\right)\,\Theta+\left[
-{m_{\psi}^2\over \cos^2\theta}-{m_{\phi}^2\over \sin^2\theta} 
+ \gamma^2 \kappa^2 \,\eta\,
\left(-n\, \sin^2\theta + (n+1)\,\cos^2\theta\right)\right]\,\Theta  = -\Lambda\,\Theta. \nn
\label{angular}
\eea where
\be
\kappa^{2}= (\omega^2-\lambda^2) a^2, 
\ee
and where $\Lambda$ is the separation constant.  
The radial equation becomes, 
\bea
&&\frac{1}{r}\frac{d}{dr}\left(r(r^2+\eta a^2  \gamma^2)\frac{dH}{dr}\right)+\Bigg\{(\omega^2- \la^2) r^2+(\omega^2-\la^2) (Q_1+Q_5)\nonumber\\
&&+(\omega-\la)^2 Q_p  +\frac{\eta a^2 }{r^2+\eta a^2  \gamma^2}\left(\frac{\omega}{\eta} R- \la R \frac{(Q_1+Q_5) Q_p}{Q_1 Q_5}-\frac{\gamma_2}{a}m_{\phi}-\frac{\gamma_1}{a}m_{\psi}\right)^2\nonumber\\
&&-\frac{a^2 \eta}{r^2}\left(\la R +\frac{\gamma_2}{a}m_{\psi}+\frac{\gamma_1}{a}m_{\phi}\right)^2-\Lambda \Bigg\}H(r)=0.
\eea
To simplify expressions, we introduce a dimensionless radial coordinate, $x$,
\be
x=\frac{r^2}{a^2}={r^2 R^2\over Q_1 Q_5}, \label{eq:x-def}
\ee
and define new parameters
\be
 \omega= {\tilde \omega \over R}, \qquad \lambda={ \tilde \lambda \over  R},
\ee
The radial equation then becomes
\bea \label{rad1}
&&4\frac{d}{dx}\left(x(x+\eta \gamma^2)\frac{dH}{dx}\right)+
\left\{(\tilde \omega^2- \tilde \lambda^2) \left[\frac{Q_1Q_5}{R^4}x+\frac{(Q_1+Q_5)}{R^2}\right] +(\tilde \omega-\tilde \la)^2 \frac{Q_p}{R^2}
\right.\nonumber\\
&&+\frac{\eta  }{x+\eta   \gamma^2}\left(\frac{\tilde \omega}{\eta} -\tilde \la  \frac{(Q_1+Q_5) Q_p}{Q_1 Q_5}-\frac{\gamma_2} {a}m_{\phi}- \frac{\gamma_1}{a}m_{\psi}\right)^2\nonumber\\
&&-\frac{ \eta}{ x}\left(\tilde \la +\frac{\gamma_2}{a} m_{\psi}+\frac{\gamma_1}{a} m_{\phi}\right)^2-\Lambda \Bigg\}H(r)=0 \ .
\eea
Defining the following convenient quantities,
 \begin{align}
&\delta~=~\frac{\sqrt{\eta}}{a} (\gamma_1+\gamma_2)~=~\sqrt{\eta}\,\gamma,  \\
& \kappa^{2} ~=~ \left[(\tilde \omega^2-\tilde \lambda^2){Q_1 Q_5\over R^4}\right], & \\
&\nu ~=~  \left(1+\Lambda - (\tilde \omega^2-\tilde \lambda^2)\frac{Q_1 + Q_5}{R^2}-(\tilde \omega-\tilde \lambda)^2 \frac{Q_p}{R^2}\right)^{1\over 2}\label{def_nu}, &\\
&\xi ~=~  \pm \sqrt{\eta}\,\left({\tilde \omega\over \eta}-\tilde \lambda\,  {Q_p (Q_1+Q_5)\over Q_1 Q_5}- \frac{m_{\psi}}{a}\,\gamma_1 - \frac{m_{\phi}}{a}\,\gamma_2 \right) \label{def_xi},\\
&\zeta ~=~ \sqrt{\eta}\,\left(\tilde \lambda + \frac{m_{\psi}}{a}\,\gamma_2 + \frac{m_{\phi}}{a}\,\gamma_1 \right), 
\label{def}
\end{align}
 the radial equation simplifies to,
\be
4{d\over d\,x}\left(x (x + \delta^2){d\over d\,x}\right) H +
\left[\kappa^{2}\,x + 1-\nu^2 +{\xi^2 \over x+\delta^2}
-{\zeta^2\over x}\right] H  =0. \label{radial_eq_full}
\ee
In the following we explore solutions to this radial equation in a matched asymptotic expansion analysis.

The scalar wave equation has a symmetry that allows one to relate solutions with positive real part of $\omega$ to negative real parts of $\omega$ \cite{Cardoso:2005gj}, i.e., if $\omega_R + i \omega_I$ is a quasinormal mode then so is $- \omega_R + i \omega_I$. This symmetry simultaneously changes the sign of $ m_\phi, m_\psi, \lambda.$ Thus without loss of generality, we can only fix the sign of one; we set $\omega_R > 0$. 
\subsection{Solutions by matching}
In order to set up a matched asymptotic expansion, we work in the parameter regime 
\be
\epsilon = \frac{(Q_1 Q_5)^\frac{1}{4}}{R} \ll 1. \label{epsilon}
\ee
In the literature this limit is often called the ``large $R$'' limit or the ``near decoupling limit''. This limit isolates the low-energy excitations of the D1-D5 bound states. In the gravity description, the geometry develops a large inner AdS$_3 \times \mathrm{S}^3$ region. Since $a = \frac{\sqrt{Q_1 Q_5}}{R}$, cf.~\eqref{parameter_a}, the limit also entails 
\be
a^2 \ll \sqrt{Q_1 Q_5}.
\ee

We will be  looking for scalar wave functions with frequency $ \omega \sim \frac{1}{R}$ and wave number  $ \lambda \sim \frac{1}{R}$. In the large $R$ limit,
in terms of $\epsilon$ defined in \eqref{epsilon}, we observe that
\bea
\kappa^2 = ( \omega^2 - \lambda^2) a^2~\sim~\epsilon^4, \qquad \eta~\sim~1, 
\eea
and so we find from the angular equation \eqref{angular} that,
\bea
\Lambda &=& l (l+2) + \mathcal{O}(\epsilon^4) \,.
\eea
We also note from eq.~\eqref{def} that 
\bea
\nu &=& l+1 + \mathcal{O}(\epsilon^2) \,.
\eea

In terms of the dimensionless radial variable $x$ defined in eq.~\eqref{eq:x-def}, we define the `inner region' to be the range  
\bea
0 \le  x \lesssim  \sigma \frac{1}{\epsilon^{2}}, \label{innerregion}
\eea
where we have introduced another parameter $\sigma \ll 1$ for convenience \cite{Chakrabarty:2015foa}.
We then define the `outer region' to be the range
\bea
x \gtrsim \frac{1}{\sigma}\frac{1}{\epsilon^{2}} \,.\label{outerregion}
\eea
As defined, the inner and outer regions do not overlap. We will match solutions in the `neck' region $x \sim \frac{1}{\epsilon^{2}}$, or more specifically,
\bea
\sigma \frac{1}{\epsilon^{2}} \lesssim x \lesssim  \frac{1}{\sigma}\frac{1}{\epsilon^{2}}.
\eea 
Solutions to the radial wave equation are power law in $x$ in this range \cite{Cvetic:1997uw}. The above definitions of inner and outer regions are valid independent of the numerical value of $l$. In particular, the discussion remains valid for $l=0$ as well. This is in contrast with the analysis of Eperon et al \cite{Eperon:2016cdd}, where a parametrically large $l$ is used to split the dimensionless radial coordinates in three regions.

\subsubsection{Solution in the inner region}

In the inner region \eqref{innerregion}, the $\kappa^2 x$ term in equation \eqref{radial_eq_full}  can be dropped. The equation then simplifies to, 
\be
\label{sol_hin}
4\frac{d}{dx}\left(x(x+k^{-2})\frac{dH(x)}{dx}\right)+\Big\{1-\nu^2+\frac{\xi^2}{x+k^{-2}}-\frac{\zeta^2}{x} \Big\}H(x)=0.
\ee
The solution of this equation that is regular at $x=0$ is of the form, 
\be
H_\rom{in}(x)=x^{\frac{k|\zeta|}{2}}  \left(x + k^{-2}\right)^{\frac{k\xi}{2}} \left[{}_2F_1(a,b; c; - x k^2)\right], \label{hinner}
\ee
where ${}_2F_1(a, b; c; z)$ is the ordinary  hypergeometric function, with
\begin{align}
a &= \frac{1}{2}\left(1 -\nu +k |\zeta| + k \xi \right), &
b &= \frac{1}{2} \left(1+ \nu + k |\zeta| + k \xi\right), &
c &= 1+  k |\zeta|. \label{def_abc}
\end{align}
In writing this solution we have chosen to normalise the wavefunction \eqref{hinner} by setting its overall normalisation constant to unity.

In the neck region, where $x \sim \frac{1}{\epsilon^2}$, the wavefunction \eqref{hinner} behaves as,
\bea 
H_\rom{in}(x) &\sim& \Gamma\left(1+  k |\zeta|\right)
\Big{[}
\frac{\Gamma(-\nu)k^{-1-\nu- k |\zeta| - k \xi}}{\Gamma\left(\frac{1}{2}(1 - \nu + k |\zeta| +k \xi)\right)\Gamma\left(\frac{1}{2}(1- \nu + k |\zeta|-  k \xi)\right)} x^{-\frac{\nu +1}{2}} \nn
& & \qquad + \frac{\Gamma(\nu)k^{-1+\nu- k |\zeta| -k  \xi } }{\Gamma\left(\frac{1}{2}(1+  \nu + k |\zeta| +  k \xi)\right)\Gamma\left(\frac{1}{2}(1+ \nu + k  |\zeta| - k \xi)\right)} x^{\frac{\nu -1}{2}} \Big{]}. 
\label{inner_expansion}
\eea
As commented above, in the neck region the scalar wave function behaves as a linear combination of two power law solutions.

\subsubsection{Outer region}
In the outer region \eqref{outerregion}, the radial equation \eqref{radial_eq_full} becomes,
\be
4x^2H''+8xH'+\Big\{\kappa^2 x+1-\nu^2 \Big\}H(x)=0.
\ee
The most general solution to this equation is a linear combination of Bessel functions,
\be \label{OuterHyperAc}
H_\rom{out}(x) = \frac{1}{\sqrt{x}}\left[ C_1 J_\nu(\kappa \sqrt{x}) + C_2 J_{-\nu}(\kappa \sqrt{x}) \right].
\ee
In the asymptotic region, $\kappa \sqrt{x} \gg 1$,  the behaviour of the outer region solution is,
\be
H_\rom{out}(x) \sim \frac{1}{x^\frac{3}{4}} \frac{1}{\sqrt{2\pi \kappa}} \left[ e^{i \kappa \sqrt{x}} e^{-i \frac{\pi}{4}}(C_1 e^{-i \nu \frac{\pi}{2}} + C_2 e^{i \nu \frac{\pi}{2}})
+  e^{-i \kappa \sqrt{x}}e^{i \frac{\pi}{4}} (C_1 e^{i \nu \frac{\pi}{2}} + C_2 e^{-i \nu \frac{\pi}{2}})\right]. \label{outer_wavefunction}
\ee
In the neck region, where   $x \sim \frac{1}{\epsilon^2}$, the combination
$\kappa \sqrt{x} \sim \epsilon^2 \cdot \frac{1}{\epsilon} \sim \epsilon \ll 1$. We use the series expansion of the Bessel function for small arguments to write
\be
H_\rom{out}(x) \sim \frac{C_1}{\Gamma(1+\nu)} \left(\frac{\kappa}{2}\right)^\nu x^{\frac{\nu-1}{2}} +   \frac{C_2}{\Gamma(1-\nu)} \left(\frac{\kappa}{2}\right)^{-\nu} x^{-\frac{\nu+1}{2}} \ .
\label{outer_expansion}
\ee

\subsubsection{Matching solutions in the neck region}
We match the asymptotic expansions \eqref{inner_expansion} and \eqref{outer_expansion} in the neck region to get
\be
\frac{C_1}{C_2}\frac{\Gamma(1-\nu)}{\Gamma(1+\nu)} \left(\frac{\kappa}{2 k}\right)^{2\nu} =   \frac{\Gamma(\nu)}{\Gamma(-\nu)}
\frac{\Gamma\left(\frac{1}{2}(1-\nu +  k |\zeta| + k \xi) \right)
\Gamma\left(\frac{1}{2}(1-\nu +  k |\zeta| - k \xi) \right)}
{\Gamma\left(\frac{1}{2}(1+\nu +  k |\zeta| + k \xi) \right)
\Gamma\left(\frac{1}{2}(1+\nu +  k |\zeta| - k \xi) \right)}.
\label{matchingnew}
\ee
We impose the no incoming waves boundary conditions at infinity, i.e., 
\be
C_1  + C_2 e^{-i \nu \pi} = 0. \label{outgoing}
\ee
Here we have assumed that $\omega_R > 0$. We obtain
\be
-e^{-i \pi \nu} \frac{\Gamma(1-\nu)}{\Gamma(1+\nu)} \left(\frac{\kappa }{2k }\right)^{2\nu} =  \frac{\Gamma(\nu)}{\Gamma(-\nu)}
\frac{\Gamma\left(\frac{1}{2}(1-\nu +  k |\zeta| + k \xi) \right)
\Gamma\left(\frac{1}{2}(1-\nu +  k |\zeta| - k \xi) \right)}
{\Gamma\left(\frac{1}{2}(1+\nu +  k |\zeta| + k \xi) \right)
\Gamma\left(\frac{1}{2}(1+\nu +  k |\zeta| - k \xi) \right)}.
\label{matchingnew2}
\ee
The emission frequencies are given by the solutions to this transcendental equation. The right hand side of this equation has $\xi$ appearing in a symmetric way under $\xi \leftrightarrow - \xi$. We can thus restrict our analysis to only one branch of the value of $\xi$, cf.~\eqref{def_xi}. We take,
\be
\xi ~=~  \sqrt{\eta}\,\left({\tilde \omega\over \eta}-\tilde \lambda\,  {Q_p (Q_1+Q_5)\over Q_1 Q_5}- \frac{m_{\psi}}{a}\,\gamma_1 - \frac{m_{\phi}}{a}\,\gamma_2 \right), \label{def_xi_new}
\ee
which in the large $R$ limit  becomes,
\be
\xi=\tilde \omega 
+ \frac{n }{k}m_{\psi}  - \frac{(n+1)}{k}m_{\phi}. \label{xi_form}
\ee
In the large $R$ limit the expression for $\zeta$ becomes
\be
\zeta = \tilde \lambda + \frac{(n+1)}{k}  m_\psi - \frac{n}{k} m_\phi. \label{zeta}
\ee

\subsection{Real and imaginary parts}
In the large $R$ limit, $\kappa$ is parametrically small and $\nu$ is close to a positive integer. Therefore,  the $\kappa^{2\nu}$ term in equation  \eqref{matchingnew2} is parametrically small. As a result, the only way the transcendental equation \eqref{matchingnew2} can be solved, is when  one of the $\Gamma$-functions  in the denominator on the right hand side is close to developing a pole. That is, to leading order, either, 
\begin{equation}
\frac{1}{2} \left(1+\nu +  k|\zeta| - k\xi \right) \simeq -N,
\label{stable_pole}
\end{equation}
with $N$ a non-negative integer, or 
\begin{equation}
\frac{1}{2} \left(1+\nu + k|\zeta| + k\xi\right) \simeq -M,
\label{unstable_pole}
\end{equation}
with $M$ a non-negative integer, with $\xi$ restricted to be of the form \eqref{xi_form}. The two solutions are,
\begin{align}
&\mbox{Modes A:}& \nn  &  \omega_R^\rom{A} \simeq \frac{1}{kR}\left[(l+2(N +1)) + |k \tilde \la+m_{\psi} (n+1)-m_{\phi} n| + (m_{\phi} (n+1) - m_{\psi} n) \right], \label{stable} & \\
&\mbox{Modes B:}& \nn &  \omega_R^\rom{B} \simeq \frac{1}{kR}\left[-(l+2(M +1)) - |k \tilde \la+m_{\psi} (n+1)-m_{\phi} n| + (m_{\phi} (n+1) - m_{\psi} n) \right] \label{unstable}.&
\end{align}
This is the complete spectrum. The subscript $R$ refers to the real part of the quasinormal mode frequencies.  The spectrum  \eqref{stable} precisely matches with expression 6.12 of \cite{Giusto:2012yz}.  In the strict decoupling limit the spectrum has the symmetry,
\be
\omega_R^\rom{A} (- \lambda, -m_{\phi},-m_{\psi}) = - \omega_R^\rom{B} (\lambda, m_{\phi},m_{\psi}).
\ee
However, in the \emph{near decoupling} limit the modes A and B are \emph{distinguished} by the imaginary parts. The imaginary parts of these quasinormal mode frequencies can be readily calculated using the procedure given in \cite{Chowdhury:2007jx, Chakrabarty:2015foa}. A simple, though somewhat lengthy, calculation gives the imaginary parts for the above modes,
\bea
\label{im_A_modes}
 \omega_I^\rom{A} &\simeq& 
- \frac{2\pi} {kR} \frac{1 }{(l!)^2}\left[(\omega_R^2- \lambda^2) \frac{Q_1 Q_5}{4 R^2 k^2}\right]^{l+1}   \  {}   ^{l+1+N} C_{l+1}  \  {}^{l+1+N+k|{\zeta}|} C_{l+1},
\\
\label{im_B_modes}
 \omega_I^\rom{B} &\simeq& + \frac{2\pi} {kR} \frac{1 }{(l!)^2}\left[(\omega_R^2- \lambda^2) \frac{Q_1 Q_5}{4 R^2 k^2}\right]^{l+1}      \ {} ^{l+1+M} C_{l+1} \  {}^{l+1+M+k|{\zeta}|} C_{l+1}, 
\eea
where ${}^nC_m$ is the standard binomial coefficient. Since $\omega_I^\rom{A}  < 0$, A modes are the stable modes, and B modes (if allowed) are the unstable modes. In writing the above expressions, implicit is the assumption that parameters can be arranged such that situation  \eqref{stable_pole} and \eqref{unstable_pole} are physically realisable. In the next subsection we argue that this assumption is false.

\subsection{No unstable modes}
\label{sec:unstable_modes}
We now argue  that the unstable modes do not exist.  Recall that we have fixed the convention $\omega_R > 0$. We now show that unstable modes do not exist by showing that we cannot make $\omega_R^\rom{B}$ and $(\omega_R^\rom{B}){} ^2 - 
\lambda^2$ positive for any value of $M, n,  k, l, m_\psi, m_\phi, \lambda$. Said differently, we cannot arrange parameters where pole \eqref{unstable_pole} is physical. Hence the procedure for computing the imaginary part following \cite{Chowdhury:2007jx, Chakrabarty:2015foa} does not go through, and the only physical modes are the stable modes. Ref.~\cite{Cardoso:2005gj} gave essentially the same argument in a slightly different form. The presentation below is instructive as almost all the emission calculations in the fuzzball literature are done in the near decoupling limit. The discussion below makes a direct connection to the CFT analysis. We will see in the next section that in the CFT emissions corresponding to B modes are forbidden.

 To begin with let us consider two charge geometries, i.e., geometries with no momentum charge, $Q_p = 0$. This happens when $n=0$ or $n=-1$.
For $n=0$, we have
\bea
 \omega_R^\rom{B} &=& \frac{1}{kR}\left[-l-2(M +1) - |k \tilde \la+m_{\psi}| + m_{\phi} \right] \\
& =&  \frac{1}{kR}\left[-(l-m_{\phi}) -2(M +1) - |k \tilde \la+m_{\psi}|  \right]. \label{2charge_B}
\eea
For spherical harmonics on the three-sphere, the values of $l, m_\phi$ and $m_\psi$ are constrained so that
\be
l \ge |m_\phi| + |m_\psi|, \qquad \qquad l \ge 0.
\ee
Thus, for $m_\phi \ge 0$, $l- m_\phi \ge 0$ and  for  $m_\phi < 0$ too  $l-m_{\phi} \ge 0$. In both cases the right hand side of equation \eqref{2charge_B} is strictly negative, i.e., such \emph{emission} modes do not exist. 
For $n=-1$, we have
\bea
 \omega_R^\rom{B} &=& \frac{1}{kR}\left[-l-2(M +1) - |k \tilde \la+m_{\phi}| + m_{\psi} \right] \\
& =&  \frac{1}{kR}\left[-(l-m_{\psi}) -2(M +1) - |k \tilde \la+m_{\phi}|  \right]. \label{2charge_B_2}
\eea
Via a similar argument (now with $m_\psi$), it follows that the right hand side of equation \eqref{2charge_B_2} is strictly negative, i.e., such modes do not exist. This is clearly in line with our expectation that the 2-charge geometries do not admit unstable modes.

For the three-charge geometries, the corresponding analysis is much more subtle.  

For the general 3-charge geometries considered above, supersymmetry property implies the existence of a causal Killing vector arising from the square of the covariantly constant Killing spinor. In 6d such a causal Killing vector is globally null~\cite{GMR}. This globally null vector does not become time translation at infinity.

The norm of the time translation Killing vector $T = \frac{\partial}{\partial t}$,
\be
T \cdot T = g_{tt} = \frac{-f + Q_p}{f h},
\ee
is positive on the ``evanescent ergosurface'' $f=0$\footnote{Evanescent ergosurface is a concept introduced in \cite{Gibbons:2013tqa} and expanded upon in \cite{Eperon:2016cdd}. In 5d it is the surface of infinite redshift relative to infinity. In 6d it is defined as the locus where globally null Killing vector field $V = \frac{\partial}{\partial t } + \frac{\partial}{\partial y}$ is orthogonal to the Kaluza-Klein Killing vector field $Y= \frac{\partial}{\partial y}$.}, where $f h = \sqrt{(f + Q_1)(f+Q_5)} = \sqrt{Q_1 Q_5}$. We have,
\be
T \cdot T  = \frac{Q_p}{\sqrt{Q_1 Q_5}} > 0, \qquad \mbox{for} \qquad Q_p > 0. 
\ee
This observation was first made in \cite{Jejjala:2005yu}.  

In fact, for every Killing vector that becomes time translation (in some boosted frame) at infinity, there is an ergoregion. The most general such vector is 
\be
\xi = \frac{\partial}{\partial t} + c \frac{\partial}{\partial y},
\ee
for $|c| < 1$.  On evanescent ergosurface $f=0$, the norm of the this Killing vector field is also positive 
\be
\xi \cdot \xi = (c-1)^2 \frac{Q_p}{\sqrt{Q_1 Q_5}} > 0.
\ee
The three-charge geometries have a genuine ergoregion in 6d.  A different perspective on this discussion is presented in~\cite{Cardoso:2007ws}.

A presence of the ergoregion signals an ergoregion instability \cite{Cardoso:2005gj}. A natural question is then: do unstable modes exist? Is it possible to arrange parameters such that $\omega_R^\rom{B} > 0$ and $(\omega_R^\rom{B}){} ^2 - 
\lambda^2 > 0$? 

We are not able to conclusively show that unstable modes do not exist, though, to the extent we have investigated, we find that it is not possible to arrange parameters such that these modes become physical.

There are a few cases that can be easily analysed analytically. For example, an inspection of the expression for $\omega_R^\rom{B}$ \eqref{unstable} shows that with fixed $n, l$ for several values on $m_\psi$ and $m_\phi$, $\omega_R^\rom{B}$ is strictly negative. All such modes are clearly unphysical. For some values of the parameters it is possible to get positive $\omega_R^\rom{B}$. For example, take $M=0, n=4, k=1, l=1, m_\phi = 0, m_\psi = -1$ and $\tilde \lambda = 5$, then 
$\tilde \omega_R = 1$. This is a potentially dangerous modes, however, since $(\tilde \omega_R^\rom{B}){} ^2 - 
\tilde \lambda^2 < 0$ such a mode is unphysical.

 To systematically search for those modes, it is convenient to introduce parameters $m$ and $\bar m$ labelling the scalar spherical harmonics basis on S$^3$, such that
\bea
m_\psi &=& \bar m - m, \\
m_\phi &=&  \bar m + m,
\eea
with $m, \bar m \in \left(-\frac{l}{2},\frac{l}{2}\right).$ For $n>0$, the values $m = \bar m = \frac{l}{2} $ and $M=0, k=1$ gives rise to the most un-favourable situation, with 
\be
\tilde \omega_R^\rom{B} = (nl -2) - | nl - \tilde \la|.
\ee
Now there are two cases: $(nl - \tilde \la) \ge 0$ and $(nl - \tilde \la)< 0$. When $(nl - \tilde \la)< 0$, we have $\tilde \lambda > nl$ and $\tilde \omega_R^\rom{B} < nl -2 $. When $(nl - \tilde \la) \ge 0$, $\tilde \omega_R^\rom{B} = \tilde \lambda - 2$ together with the condition $\tilde \la \ge 2$. In both these siutations $(\tilde \omega_R^\rom{B}){} ^2 - 
\tilde \lambda^2 < 0$, i.e., such modes are unphysical. 

More cases can be considered along the similar lines. We have done an extensive computer  search. We organised our computer program as follows: for fixed $n, N, l$ we find the minimum and maximum values of $\lambda$ for which $\omega_R^\rom{B}$ is positive by varying $m$ and $\bar m$. Then we check for the condition $(\omega_R^\rom{B}){} ^2 - \lambda^2 < 0$.  We do not find any unstable mode.\footnote{We thank Samir Mathur for several useful discussions on these issues.} Writing a full combinatorial proof that no such modes exist is likely to be cumbersome, involving many cases. In the absence of such a proof, it remains a conjecture that one cannot arrange parameters so that unstable modes become physical. 

For simplicity of notation from now onwards we will drop the superscripts A and B and exclusively work with the stable modes (A modes).

\subsection{Eperon-Reall-Santos limit}
\label{ERS_limit}
For black holes, quasinormal modes in the large $l$ limit can be related to properties of circular null geodesics  \cite{Ferrari:1984zz, Cardoso:2008bp, Yang:2012he}.  Eperon, Reall, and Santos (ERS) \cite{Eperon:2016cdd} analysed the quasinormal mode spectrum in the large $l$  limit for the supersymmetric microstate geometries.  They focus on two classes of modes: (i) modes for which the corresponding (stably) trapped null geodesics are at the evanescent ergosurface, i.e., the energy and Kaluza-Klein momentum of these modes do not scale with $l$, (ii) modes for which the corresponding (stably) trapped null geodesics are \emph{away} from the evanescent ergosurface, i.e., the energy and Kaluza-Klein momentum of these modes scale with $l$ in such a way that $\tilde \omega_R + \tilde \lambda \sim \mathcal{O}(l)$ but $\tilde \omega_R - \tilde \lambda \sim \mathcal{O}(l^0)$. In the large $l$ limit (geometrical optics  limit) \emph{and} the large $R$ limit (near decoupling limit) several of their expressions can be recovered from ours. For this discussion we keep the orbifolding parameter $k$ unfixed; to match with the ERS expressions one needs to set $k=1$.

\subsubsection{$ k \tilde \omega_R \ll l$ and $k |\tilde \lambda| \ll l$: modes at evanescent ergosurface}

\paragraph{Real part:} We start by looking at the real parts of the frequencies. 

\paragraph{\underline{For $n>0$:}}

In order to achieve  $ k \tilde \omega_R \ll l$ with $k |\tilde \lambda| \ll l$, the terms proportional to $l$ should cancel out in $\tilde \omega_R$. This is possible~\cite{Eperon:2016cdd} for $n>0$ with
\be
m_\phi + m_\psi = -l, 
\ee
and
\be
0 \le \frac{m_\psi}{m_\phi} \le \frac{n}{n+1}.
\ee
From these relations it follows that both $m_\phi$ and $m_\psi$ are negative. Therefore, for these modes, 
\be
 |m_{\psi} (n+1)-m_{\phi} n| =  m_{\psi} (n+1) -m_{\phi} n.
\ee
To leading order in $l$ we arrive at,
\bea
k \tilde  \omega_R &=& l + m_{\psi} (n+1) -m_{\phi} n + (m_{\phi} (n+1) - m_{\psi} n) + k |\tilde \lambda| + 2 (N+1) +\mathcal{O}(l^{-1})  \\ &=&k|\tilde \lambda| + 2 (N+1) +\mathcal{O}(l^{-1}). \label{ERS_85}
\eea
This last equation is equation (85) of ERS. 

\paragraph{\underline{For $n<-1$:}} Terms proportional to $l$ cancel out in $\tilde \omega_R$ with
\be
m_\phi + m_\psi = -l, 
\ee
and
\be
\frac{m_\psi}{m_\phi} \ge \frac{n}{n+1}.
\ee
From these relations again it follows that both $m_\phi$ and $m_\psi$ are negative. Therefore, for these modes, 
\be
 |m_{\psi} (n+1)-m_{\phi} n| =  m_{\psi} (n+1) -m_{\phi} n.
\ee
We get,
\bea
k \tilde \omega_R  &=& l + m_{\psi} (n+1) -m_{\phi} n + (m_{\phi} (n+1) - m_{\psi} n) + k|\tilde \lambda| + 2 (N+1) +\mathcal{O}(l^{-1})  \\ &=&k|\tilde \lambda| + 2 (N+1) +\mathcal{O}(l^{-1}).
\eea
This last equation is again equation (85) of ERS.

\paragraph{\underline{For $n=0$:}} This case corresponds to a 2-charge geometry with $Q_p = 0$. For this case, the ERS modes are with
\be
m_\phi = -l, \qquad m_\psi = 0.
\ee
For these modes, 
\bea
k \tilde \omega_R &=& k|\tilde \lambda| + 2 (N+1) +\mathcal{O}(l^{-1}).
\eea
This is equation (114) of ERS.

 \paragraph{\underline{For $n=-1$:}} This case also corresponds to a 2-charge geometry  -- a different 2-charge geometry than the $n=0$ case. For this case, ERS modes are with,
\be
m_\phi = 0, \qquad m_\psi = -l.
\ee
For these modes too, 
\bea
 k \tilde \omega_R  &=& k|\tilde \lambda| + 2 (N+1) +\mathcal{O}(l^{-1}).
\eea
This last equation is again equation (114) of ERS. 

\paragraph{Imaginary part:} Now let us look at the imaginary part of the frequencies.  Let 
\be
j \equiv \frac{m_\psi}{m_\phi}, \qquad m \equiv \frac{m_\phi}{l}.
\ee 
In the large $l$ limit, for modes with $k|\tilde \lambda| \ll l$ we have
\bea
k |{\zeta}| &=& |k \tilde \lambda + m_{\psi} (n+1)-m_{\phi} n| =  m_{\psi} (n+1) -m_{\phi} n + \mathcal{O}(l^0) \\
&=&  m_{\phi}  \left(j(n+1) -n\right) +  \mathcal{O}(l^0) \\
&=&  m l   \left(j(n+1) -n\right) +  \mathcal{O}(l^0) =\mu  l  +  \mathcal{O}(l^0) \label{def_mu}
\eea
where
\be
\mu = - \frac{\left(j(n+1) -n\right)}{1+j} > 0.
\ee
Using Stirling's approximation in various factors of eq.~\eqref{im_A_modes} with the value \eqref{def_mu} of $k |\zeta|$, we find 
\bea
\frac{1}{(l!)^2} &\sim&  \exp\left[ - 2 l \ln l + 2 l - \ln l  + \mathcal{O}(l^0) \right], \label{Stirling_1}\\
 {}   ^{l+1+N} C_{l+1} & \sim&  \exp \left[ N \ln l  + \mathcal{O}(l^0)\right],  \label{Stirling_2} \\
{}^{l+1+N+k|{\zeta}|} C_{l+1} &\sim&  \exp\left[ \left\{ -\mu \ln \mu + (1 +  \mu)\ln (1 + \mu) \right\} l- \frac{1}{2}\ln l + \mathcal{O}(l^0)\right]. \label{Stirling_3}
\eea
The remaining dimensionless factor in eq.~\eqref{im_A_modes} behaves as, 
\be
\left[ ( \omega^2-\lambda^2) \frac{Q_1 Q_5}{4 R^2 k^2}\right]^{l+1}  = \left[\frac{\kappa^2}{4 k^2} \right]^{l+1} ~\sim~ \kappa_0^2 \exp \left[ l \ln \left[ \frac{\kappa_0^2}{4k^2} \right] + \mathcal{O}(l^0)\right], \\
\ee
where $\kappa^2 = (\omega^2-\lambda^2)a^2$ and $\kappa_0^2 = (\omega_R^2- \lambda^2)a^2$. Multiplying these factors, we obtain the imaginary part of the frequency in the large $l$ limit. The resulting expression  matches with the corresponding ERS expression, eq.~(89), in the decoupling limit (with $k=1$),
\be
\tilde \omega_I \sim  - D \kappa_0^2 e^{- 2 \, l \ln l+  \left\{(2  -\mu \ln \mu + (1 +  \mu)\ln (1 + \mu) +  \ln \frac{\kappa_0^2}{4k^2} \right\} l + \left( N - \frac{3}{2} \right) \ln l + \mathcal{O}(l^0)} ,
\ee
where $D$ is a positive constant independent of $l$ to the leading order in $l$.\footnote{An expression for the coefficient $D$ can certainly be obtained, but it is somewhat cumbersome as it depends on $\mathcal{O}(l^0) $ correction to  $|\zeta|$ through equation \eqref{def_mu}.}

\subsubsection{$ k ( \tilde \omega_R +\tilde \lambda)  \sim \mathcal{O}(l)$ and $k (\tilde \omega_R - \tilde \lambda)  \ll l $: modes away from evanescent ergosurface}
For $n > 0$, ERS also looked at modes with $\lambda > 0$ and with
\be
m_\phi + m_\psi = -l, 
\ee
and
\be
0 \le \frac{m_\psi}{m_\phi} \le \frac{n}{n+1}.
\ee
 For such modes, to leading order in $l$ we have,
\bea
k \tilde \omega_R  &=& l + m_{\psi} (n+1) -m_{\phi} n + (m_{\phi} (n+1) - m_{\psi} n) + k \tilde \lambda + 2 (N+1) +\mathcal{O}(l^{-1})  \\ &=&k\tilde \lambda + 2 (N+1) +\mathcal{O}(l^{-1}). \label{ERS_94}
\eea
This expression is same as the above expression \eqref{ERS_85}, except now $k \tilde \lambda$ scales with $l$. In the decoupling limit this expression matches with eq.~(94) of ERS. The $1/R$ corrections to \eqref{ERS_94} obtained in \cite{Eperon:2016cdd} cannot be captured by our analysis as we have already taken the large $R$ limit.  For $n< -1$ the corresponding expressions are readily obtained from the analysis of the previous subsection.  Once again the real part of the quasinormal mode frequencies is \eqref{ERS_94}. 

\paragraph{Imaginary part:} For the imaginary part, we start with the observation 
\bea
k |{\zeta}| &=& |k \tilde \lambda + m_{\psi} (n+1)-m_{\phi} n| =  \tilde \lambda + m_{\psi} (n+1) -m_{\phi} n + \mathcal{O}(l^0) \\
&=&  k\tilde \lambda + m_{\phi}  \left(j(n+1) -n\right) +  \mathcal{O}(l^0) \\
&=& k\tilde \lambda +  m l   \left(j(n+1) -n\right) +  \mathcal{O}(l^0) =\mu'  l  +  \mathcal{O}(l^0)
\eea
where
\be
\mu' = \frac{k \tilde \lambda}{l}- \frac{\left(j(n+1) -n\right)}{1+j}.
\ee
Let $\mu'  > 0 $. Using Stirling's approximation \eqref{Stirling_1}--\eqref{Stirling_3} in various factors of eq.~\eqref{im_A_modes} together with
\be
(\tilde{\omega}^2-\tilde{\lambda}^2)  = (\tilde\omega - \tilde\lambda) (\tilde\omega + \tilde\lambda) \sim l, 
\ee
and
\bea
\left[ (\omega^2- \lambda^2) \frac{Q_1 Q_5}{4 R^2 k^2}\right]^{l+1}  
& \sim&
 \exp \left[l \ln l + l  \ln \left[ \frac{\kappa_0^2}{4l k^2} \right] + \ln l+ \mathcal{O}(l^0)\right],
\eea
we obtain the large $l$ limit of the imaginary part of the quasinormal mode frequencies. 
Multiplying various factors, we obtain
\be
\tilde \omega_I^\rom{A} \sim  - D'  e^{-   \, l \ln l+  \left\{(2  -\mu' \ln  \mu' + (1 +  \mu')\ln (1 + \mu') +  \ln \frac{\kappa_0^2}{4 l k^2} \right\} l + \left( N - \frac{1}{2} \right) \ln l + \mathcal{O}(l^0)} ,
\ee
where $D'$ is a positive constant independent of $l$ to the leading order in $l$. It is proportional to $\frac{Q_1 Q_5}{4 R^2 k^2}.$ This expression  matches with the corresponding ERS expression, eq.~(96), in the decoupling limit (with $k=1$).

To summarise: most of the interesting features of the ERS modes are captured in our  near decoupling limit analysis.

\section{Quasinormal Modes from a D1-D5 orbifold CFT analysis}
\label{sec:CFT}
The aim of this section is to obtain the real and imaginary parts of the quasinormal mode spectrum computed in the previous section from  a CFT analysis. This involves several key steps. Fortunately, almost all of these steps are already carefully presented in \cite{Avery:2009tu}; see also earlier papers \cite{Chowdhury:2007jx, Chowdhury:2008bd, Chowdhury:2008uj} and the review \cite{Chowdhury:2010ct}. We will see that a correct interpretation of some of intermediate results from these references gives the full spectrum. To keep the discussion simple we set $k=1$. Towards the end we briefly indicate how the discussion generalises to $k>1$.  We wish to emphasise that although the correct interpretation of some of the known results is all that is required,  the points made below are not well appreciated in the literature. To the best of our knowledge no study exists that tries to make the connection between the \emph{stable} quasinormal modes of any D1-D5 microstate geometry and the D1-D5 CFT.  
 
\subsection{Scalar emission from the D1-D5 CFT  }

To set the notation we start by describing some elementary features of the D1-D5 CFT.  Consider type IIB string theory compactified on $S^{1} \times T^{4}$ with $N_{1}$ number of D1 branes wrapping $S^1$ and $N_{5}$ number of D5 branes wrapping $S^1\times T^{4}$.  The bound state of these branes can be described by a field theory. We take the size of the torus to be of the order of the string scale whereas we take the radius $R$ of the circle $S^1$ to be large compared to the string scale. For such a set up, at low energy, we can focus on the excitations only on the $S^1$ direction. This low energy limit gives a two-dimensional CFT on the circle $S^1$. It is conjectured that we can move  in the moduli space of this CFT to the `orbifold point' where the CFT is a $\mathcal{N}=(4,4)$ supersymmetric $(1+1)$ dimensional sigma model whose target space is the symmetrized product of $N_{1}N_{5}$ copies of $T^4$; for more details and references see e.g.,~\cite{David:2002wn, Avery:2010qw}. A convenient way of  thinking about the sigma model is in terms of component strings. Each copy of the $T^4$ is viewed as a component string, giving 4 bosonic excitations together with 4 left moving and 4 right moving fermionic excitations.  Since the sigma model target space is the symmetrized product, there are twisted sectors. The twisted sector states can be obtained by applying twist operators $\sigma_n$ on an untwisted state. The twist operators link several copies of the component strings~\cite{Lunin:2000yv}.

The CFT that describes the bound state of the D1-D5 system is in the Ramond sector. The Ramond sector has  a number of degenerate ground states. All these ground states can be obtained by applying `one unit' of spectral flow on chiral primary states in the NS sector. The gravity dual of the NS vacuum state $\vac_{NS}$ is $AdS_{3}\times S^{3}\times T^{4}$. One unit of spectral flow on the left and right sectors on the NS vacuum gives a Ramond sector ground state. Further even units of spectral flows only on the left sector give supersymmetric excited states. The GMS microstate geometries are dual to precisely these state. To summarise: the three-charge GMS geometries are dual to odd units of spectral flows on the left and one unit of spectral flow on the right on the NS vacuum. Due to the fact that the GMS geometries have a direct relation to the NS vacuum,  several calculations can be first done in the NS-NS sector and then spectral flowed to the Ramond-Ramond sector.

The  $\mathcal{N}=4$ superconformal symmetry of the D1-D5 CFT is generated by $L_n,  G^{\pm}_r, J_n^a$ on the left and $\bar L_n,  \bar  G^{\pm}_r, \bar  J_n^a$ on the right. Expressions for the left and right generators in terms of the free fields can be found in e.g.,~\cite{Avery:2009tu, Avery:2010qw} together with relevant OPEs and the (anti-)commutation relations. 
The generators  $J_0^a$ and $\bar J_0^a$ correspond to the $\mathrm{SU}(2)_L \times \mathrm{SU}(2)_R$ R-symmetry of the D1-D5 CFT.  We denote the quantum numbers in the left and the right SU(2) as $(j,m)$ and $(\bar j, \bar m)$ respectively. The quantum number under $L_0$ and $\bar L_0$ are denoted $h$ and $\bar h$ respectively. In addition, there is an internal symmetry corresponding to the rotation on the $T^4$: $\mathrm{SO}(4)_I \simeq \mathrm{SU}(2)_1 \times  \mathrm{SU}(2)_2$  . This symmetry is broken by the compactification of the torus, but it is useful in organising the field content of the D1-D5 CFT at the orbifold point. The indices $A, B$ are used for the doublet of the SU(2)$_1$ and the indices $\dot A, \dot B$ are used for the doublet of the SU(2)$_2$. The torus indices $i$ and $j$ are related to the $A, \dot A$ indices via the Pauli matrices. For a vector $X^i$ on the internal torus $T^4$,  we have
\be
[X]{}^{ \dot A A}= \sum_{i=1}^{4}X^i (\sigma^i){}^{\dot A A},
\ee
where $\sigma^{i}$ are the three Pauli matrices (for $i=1,2,3$) and the identity matrix ($\sigma^4 = i \mathbb{I}_2$). The notation is more fully explained in references \cite{Avery:2009tu, Avery:2010qw}.

We are interested in studying minimally coupled massless scalar in the GMS geometries. The minimally coupled scalar in six-dimensions correspond to the ten-dimensional graviton with indices in the torus directions. From the study of the spectrum of IIB supergravity on $AdS_3 \times S^3 \times T^4$, it follows that  in the NS sector the scalar excitations can be described as descendants of chiral primary states; see e.g. discussion in~\cite{David:2002wn}. Thus we have the following picture: adding an excitation in the AdS region of the geometry corresponds to an excitation in the CFT. This excitation relaxes by emitting a scalar quanta.  The initial state is related via spectral flows to descendants of chiral primary states. The final state is the state dual to the three-charge GMS geometry. We describe our initial and final state and the vertex operator following in more detail in the next subsection.

The emission is caused by the coupling of the CFT to modes to infinity. The general structure of such coupling was discussed in \cite{Avery:2009tu}. Our analysis heavily relies on those results. The idea is to compute a transition amplitude in the CFT (which in the present set-up is essentially a two-point function) and use it to compute the real and imaginary parts of the quasinormal mode spectrum. A related analysis for AdS$_5$ black holes was reported in \cite{Rocha:2008fe}.

\subsection{Initial and final states}
We consider the scenario where the initial state is an excited state. It is dual to the three-charge geometry \emph{with} the scalar excitation.  This excited state decays to a lower energy state. The lower energy state is dual to an unexcited three-charge GMS geometry. In this transition, a scalar quanta is emitted. As mentioned above, since the GMS geometries have a direct relation to the NS vacuum,  the calculation of transition amplitude can be first done in the NS-NS sector and then spectral flowed to the Ramond-Ramond sector. The final state is the NS description is simply the NS-NS vacuum,
\be
|f\rangle = \vac_\rom{NS}.
\ee
The correctly normalised NS-NS sector excited state was constructed in~\cite{Avery:2009tu}. It takes the form, 
\begin{align}
|i\rangle = \ket{\phi}^{A\dot{A}B\dot{B}} 
      &= \sqrt{\frac{(l-q)!(l-\bar{q})!}{4N!\bar{N}!(N+l+1)!(\bar{N}+l+1)!q!\bar{q}!(l+1)^2}}\nonumber\\
	&\qquad\times L_{-1}^N(J_0^-)^q  G^{-A}_{-\frac{1}{2}}
	\psi_{-\frac{1}{2}}^{+\dot{A}}\,\,
	\bar{L}_{-1}^{\bar{N}}(\bar{J}_0^-)^{\bar{q}} \bar{G}^{\dot{-}B}_{-\frac{1}{2}}
	\bar{\psi}_{-\frac{1}{2}}^{\dot{+}\dot{B}}\sigma^0_{l+1}\vac_\rom{NS}. \label{excited_NS}
\end{align}
The notation is as follows: 
\begin{enumerate}
\item In equation \eqref{excited_NS}, the operator $\sigma^0_{l+1}$ is the normalised chiral primary operator with dimensions (under $L_0, \bar{L}_0$) and SU(2) R-charges (under $J^3_0, \bar{J}^3_0$),
\begin{align}
\sigma_{l+1}^0 : & & h = m = \frac{l}{2}, &  &  \bar h = \bar m = \frac{l}{2}.
\end{align} 
We will also need the anti-chiral primary operator $\tilde \sigma^0_{l+1}$. It has dimensions and charges,
\begin{align}
\tilde \sigma_{l+1}^0: & & h = -m = \frac{l}{2},  & &   \bar h = -\bar m = \frac{l}{2}.
\end{align} 
The normalisation of these operators is such that the two-point function is unit normalised,
\be
\langle \tilde \sigma^0_{l+1}(z) \sigma^0_{l+1} (0)\rangle = \frac{1}{|z|^{l}}. \label{twist_correlation_func}
\ee
\item The $J_0^-$ and $\bar{J}_0^-$ are respectively the lowering operators for the left and right SU(2) R-charges. They act in equation \eqref{excited_NS} $q$ and $\bar{q}$ times respectively.  The SU(2) angular momentum quantum number of the initial state \eqref{excited_NS} are therefore,
\begin{align}
m &=\frac{l}{2} - q, & \bar m &= \frac{l}{2} - \bar q, 
\end{align} 
which are related to the SO(4) quantum numbers (in our notation) as,
\begin{align}
 m_\phi &\equiv \bar m +  m = l - q - \bar q, &m_\psi &\equiv \bar m -  m = q - \bar q. \label{m_psi_m_phi}
\end{align} 
These last relations are different from what is used in reference~\cite{Avery:2009tu}. A reason for this difference is that they work with three-charge non-supersymmetric geometries, which are often written in different angular coordinate conventions. Our notation is same as that of \cite{Giusto:2012yz}. The $m_\phi$ and $m_\psi$ are the angular momentum quantum numbers of the emitted scalar,~cf.~\eqref{scalar_ansatz}.

\item The action of  $L_{-1}^N$ and $\bar{L}_{-1}^{\bar{N}}$ increases and energy of the state by $N$ and $\bar{N}$ respectively on the left and the right sectors. These integers account for the `harmonics' in the quasinormal mode spectrum, that is the integers $N$ and $M$ that feature in equations~\eqref{stable}--\eqref{unstable}.

\item Since the scalar field is viewed as a 10d graviton with legs in the $T^4$ directions, the state \eqref{excited_NS}  carries ${A\dot{A}B\dot{B}}$ as its free indices.
 \end{enumerate}

\subsection{Scalar emission vertex operator}
The scalar emission vertex operator is obtained by starting with a twist operator in the CFT and dressing it appropriately. This operator was  constructed in detail in \cite{Avery:2009tu}. In the notation of appendix A of that reference the correctly normalised operator which corresponds to the \emph{scalar emission with} SO(4) charges $l, m_\phi, m_\psi$ is (note that the vertex operator itself has opposite SO(4) charges),
\be
\cV^{A \dot A B \dot B}_{l, m_\phi, m_\psi} (z, \bar z) = \frac{1}{2} \sqrt{\frac{(l-q)! (l-\bar q)!}{(l+1)^2 (l+1)!^2 q! \bar q!}} \left( (J_0^+)^q  (\bar J_0^+)^{\bar q} G_{-\frac{1}{2}}^{+A}
\psi_{-\frac{1}{2}}^{- \dot A}
\bar G_{-\frac{1}{2}}^{\dot +  B}
\psi_{-\frac{1}{2}}^{ \dot - \dot B}
\tilde \sigma_{l+1}^0 (z,\bar z) \right)_{z,\bar z}.
\ee

Note that since $\tilde \sigma_{l+1}^0$ appears in the vertex operator and the $\sigma_{l+1}^0$ appears in the initial state, the computation of the emission amplitude $\langle i | \cV | f \rangle$ essentially becomes a two-point function calculation for the twist operators, cf.~\eqref{twist_correlation_func}. 

\subsection{Real part of the frequency}
Now we  can put together various pieces and get the spectrum and the emission rates from the excited CFT state.  We are interested in a process involving a single quanta.  Precisely this computation was done in \cite{Avery:2009tu} as an intermediate step; the result can be read-off from that paper, equation (8.4). Translated into our conventions\footnote{Apart from the differences mentioned around equation \eqref{m_psi_m_phi}, the only other difference is that in our convention $P_y = L_0 - \bar L_0 = - P_y^\rom{there}. $}, we have for the scalar emission spectrum from the CFT side to be, 
\bea
\omega_R &=& \frac{1}{R} \left[(\alpha + \bar{\alpha} + 2)\tfrac{l}{2} - \alpha q - \bar{\alpha}\bar{q}
	+ N + \bar{N} + 2\right], \label{energy_CFT} \\
\lambda &=& \frac{1}{R}\left[(\alpha -\bar{\alpha})\tfrac{l}{2} - \alpha q + \bar{\alpha}\bar{q}
	+ N - \bar{N}\right]. \label{lambda_CFT}
\eea
In this equation the parameters $\alpha$ and $\bar \alpha$ are the spectral flows on the left and the right sectors respectively. We reach the three-charge states of interest from the NS-NS vacuum by applying $\alpha=(2n+1)$ units of spectral flows on the left and applying $ \bar{\alpha}=1$  unit of spectral flow on the right. 

The above spectrum is written slightly differently than on the gravity side. On the gravity side one only integer features and $ \lambda$ is thought of an as independent parameter. We can match the two answers as follows.  Solving \eqref{m_psi_m_phi} for $q$ and $\bar q$ in favour of $m_\psi$ and $m_\phi$, we get
\bea
q&=&\frac{1}{2} (l+m_{\psi}-m_{\phi}), \\
\bar{q}&=&\frac{1}{2} (l-m_{\psi}-m_{\phi}). 
\eea
Substituting $\alpha, \bar{\alpha}, q, \bar{q}$ in \eqref{lambda_CFT} we get
\be
\tilde \lambda =  N - \bar N - m_{\psi} (n+1) + m_{\phi} n.
\ee
Rewriting it in terms of $\zeta$,~cf.\eqref{zeta}, we have $N  - \bar{N} = \zeta$. For $\zeta \ge 0$, substituting $N = \bar N + \zeta$ we get,
\be
 \omega_R =  \frac{1}{R}\left[(l+2(\bar N +1)) +\zeta+ (m_{\phi} (n+1) - m_{\psi} n) \right],
\ee 
and for $\zeta < 0$, substituting $ \bar N =  N + |\zeta|$ we get,
\be
 \omega_R = \frac{1}{R}\left[(l+2(N +1)) + |\zeta|+ (m_{\phi} (n+1) - m_{\psi} n) \right].
\ee
For both cases combined we can write the $\omega_R$ expression as 
\be
 \omega_R = \frac{1}{R}\left[(l+2(M +1)) + |\tilde \lambda+m_{\psi} (n+1)-m_{\phi} n|+ (m_{\phi} (n+1) - m_{\psi} n) \right],
\ee
for some positive integer $M$. This expression is precisely equation~\eqref{stable} for $k=1$. There are no other modes. 

This discussion can be readily generalised to $k > 1$. The relevant equations can now be obtained as intermediate steps from reference~\cite{Avery:2009xr}. In that reference the authors calculate in the CFT an amplitude for the twisting process as they are interested in reproducing the unstable spectrum of the non-extremal JMaRT geometries. In this paper we are interested in corresponding CFT amplitude for the \emph{untwisting} process.  The answer for the energy spectrum in our conventions is 
\bea
\omega_R &=& \frac{1}{kR} \left[(k\alpha + k\bar{\alpha} + 2)\tfrac{l}{2} -k \alpha q - k \bar{\alpha}\bar{q}
	+ N + \bar{N} + 2\right], \label{energy_CFT} \\
\lambda &=& \frac{1}{R}\left[(k\alpha -k\bar{\alpha})\tfrac{l}{2} - k \alpha q + k \bar{\alpha}\bar{q}
	+ N - \bar{N}\right]. \label{lambda_CFT}
\eea
Following exactly the same logic as above, one sees that the resulting expression matches with the gravity answer for $k>1$ too,  equation \eqref{stable}. See also discussion in section 6.2 of \cite{Giusto:2012yz}.

\subsection{Imaginary part of the frequency}
Typically in AdS/CFT one computes correlation functions in the CFT and compares them to quantities computed in the AdS geometry. The set-up we are interested in is slightly different. The three-charge microstate geometries discussed above have an inner AdS region glued to an asymptotically flat region. The quansinormal modes are determined via the outgoing boundary conditions in the asymptotically flat region. Therefore, we are interested in the emission of a scalar quanta leaving to infinity of the asymptotically flat region. This requires coupling of CFT to flat space modes. This physics was worked out in detail in reference~\cite{Avery:2009tu}; see also earlier paper \cite{Chowdhury:2007jx}. The basic ideas we describe in words and refer the reader to those references for details.

 AdS/CFT allows us to relate the partition function for gravity in AdS with given field values at the boundary of AdS to the CFT partition function with sources. The boundary values of the fields $ \phi_b(y)$ act as sources $J(y)$ for the operators $\mathcal{O}(y)$ dual to the field, 
\be
S_\rom{int} = - \int d^2y J(y) \mathcal{O}(y) = - \mu \int d^2y \phi_b(y) \mathcal{V}(y), \label{coupling}
\ee
where $d^2y$ is the two dimensional $(t,y)$ space on which the CFT lives,  $\phi_b(y)$ is boundary value of the field $\phi(x)$ and $\mathcal{V}(y)$ is the operator dual to the scalar field. The scalar field and the the operator $\mathcal{V}(y)$  are appropriately normalised. Using AdS/CFT prescription we can compute the two-point function of the operator $\mu \mathcal{V}(y)$. Using the chosen normalisation of $\mathcal{V}(y)$ we can fix the coupling constant $\mu$. 

The boundary value of the field $\phi_b(y)$ can also be viewed as the limiting value of the field from the outer region. This  allows to continue modes from the AdS region to the asymptotically flat region. Suppose we have an excited initial state in the inner AdS region $|i\rangle$ and vacuum in the outer asymptotically flat region $\vac_\rom{outer}$. Due to the coupling \eqref{coupling} a particle can be emitted to infinity, lowering the energy and the other quantum numbers of the inner region state. Let the final inner region state be $|f\rangle$. The final outer region state be a one-particle state. The total amplitude for the process (to the lowest order approximation in the coupling) is 
\be
\bigg[ \langle 1_\rom{particle} | \times \langle f | \bigg]  (i S_\rom{int}) \bigg[ |i\rangle \times \vac_\rom{outer} \bigg].
\ee
From such considerations, an expression for the rate of radiation to infinity in terms of the CFT amplitude for the decay of an excited state was written in \cite{Avery:2009tu}.  The decay rate takes the form equation~(2.65) of that reference,
\be
\frac{\mathrm{d} \Gamma}{\mathrm{d} E} = \frac{2\pi}{2^{2l+1} (l!)^2} \frac{(Q_1 Q_5)^{l+1}}{R^{2l+3}} ( \omega^2 -  \lambda^2)^{l+1} |\langle f |  \cV |i \rangle_\rom{unit}|^2 \delta_{ \lambda,  \lambda_0} \delta( \omega- \omega_0). \label{rate}
\ee

In writing the above expression we have in mind the emission of a minimally coupled massless scalar with angular momentum mode $l$ on the $S^3$ from the D1-D5 system. The ansatz  for the scalar wavefunction is \eqref{scalar_ansatz};  $R$ being the radius of the $S^1$ at infinity; $ \omega_0$, $\lambda_0$ and the angular quantum numbers of the emitted quanta are determined by the difference in the quantum numbers of the initial and final CFT states; $\langle f |  \cV |i \rangle_\rom{unit}$ is the CFT correlator on a `unit sized' cylinder of cirumference $2\pi$.  The
decay rate \eqref{rate} refers to the  decay of the intensity of the radiation, whereas $\omega^\rom{A}_I$ refers to the decay of the amplitude of the wave. Since the intensity is quadratic in the amplitude, $\omega^\rom{A}_I$ is half of the $\frac{\mathrm{d} \Gamma}{\mathrm{d} E} $. Moreover, since $\frac{\mathrm{d} \Gamma}{\mathrm{d} E} $ refers to the decay, the correct interpretation in terms of the quasinormal modes is with the minus sign,
\be
\omega^\rom{A}_I = - \frac{1}{2}\frac{\mathrm{d} \Gamma}{\mathrm{d} E}.
\ee

As an intermediate step of the analysis in \cite{Avery:2009tu} the amplitude $\langle f |  \cV |i \rangle_\rom{unit}$ for precisely the process of current interest was computed. The final emission rate can be read-off from equation (8.6) of that reference. Appropriate generalisation for $k> 1$ via reference \cite{Avery:2009xr} gives the answer, 
 \bea
 \omega_I^\rom{A} =
- \frac{2\pi} {kR} \frac{1 }{(l!)^2}\left[(\omega^2- \lambda^2) \frac{Q_1 Q_5}{4 R^2 k^2}\right]^{l+1}   \  {}   ^{l+1+N} C_{l+1}  \  {}^{l+1+N+k|{\zeta}|} C_{l+1}.
\eea
This expression matches with the gravity calculation \eqref{im_A_modes}.

To summarise: a correct interpretation of the known results in the context of the D1-D5 system allow to capture the full scalar quasinormal mode spectrum from the D1-D5 CFT in the near decoupling limit.

As we saw in detail in section \eqref{ERS_limit}, this spectrum includes the slow decaying ERS modes.  

\section{Analysing the scalar wavefunction}
\label{sec:wave_func}
We now analyse the scalar wavefunction and certain features of the emitted scalar radiation. We investigate the physical picture that the decay process corresponds to the leakage of excitation from AdS throat to infinity.

Our considerations are inspired by the analysis of \cite{Chowdhury:2008bd, Chakrabarty:2015foa}, though the picture and the set-up are completely different.  In those references \emph{unstable} modes of the non-extremal JMaRT solution were analysed. It was argued that the emission process corresponds to a pair creation. The positive energy excitation escapes to infinity and a negative energy excitation settles down in the ergoregion. It was shown there that the inner region and the outer region excitations carry equal and \emph{opposite} values of various charges. As we will see below shortly, that this is not the relevant picture for our case. In the present context the decay process is as in the $\alpha$-decay from an atomic nucleus.    Scalar excitation slowly leaks from the inner region (AdS throat) to the outer region (asymptotic infinity). 

 Let $J^\mu$ be a conserved current for the perturbation, $\nabla_\mu J^\mu = 0$.  Since our background configuration has four Killing vectors, namely $\partial_t, \partial_y, \partial_\phi,$ and $\partial_\psi$, we can construct four such conserved currents for the scalar perturbation with ansatz \eqref{scalar_ansatz}. 
The energy momentum tensor of the (complex) scalar field $T_\mu{}^\nu$ is 
\be
T_{\mu \nu} = \partial_{\mu} \Phi \partial_{\nu} \Phi^* + \partial_{\nu} \Phi \partial_{\mu} \Phi^* - g_{\mu \nu} \partial_\alpha \Phi \partial^\alpha \Phi^*. \label{stress_tensor}
\ee
Conserved currents are simply $T_t{}^{\mu}, T_y{}^{\mu}, T_\psi{}^{\mu},$ and $T_\phi{}^{\mu}$.

With the scalar field ansatz \eqref{scalar_ansatz}, we saw above that the imaginary parts of the quasinormal mode frequencies are negative. As a result, near infinity the scalar solutions behave as (for the simple case of $\lambda = 0$),
\be
\Phi \sim \exp \left[ -  i \omega (t-r)\right]  \sim \exp \left[  \omega_I t - \omega_I r \right].
\ee
For $\omega_I < 0$, solutions decay in time (as expected), but grow exponentially at large distances.  This growth at infinity is a well known feature of the stable quasinormal modes, see e.g.~discussion and references in section 3.1 of \cite{Berti:2009kk}.\footnote{In general quasinormal modes can only  be thought of as quasi-stationary states that cannot have existed for all times. They are excited at a particular instant in time in a local region of the spacetime and they decay exponentially with time.} Due to this exponentially growing nature of the modes at large distances,  we cannot follow the discussion in references \cite{Chowdhury:2008bd, Chakrabarty:2015foa} to construct conserved quantities 
\be
Q_\rom{total} = \int_{\Sigma} J^\nu dS_v,
\ee
where the integral extends over a complete spacelike hypersurface in the spacetime. In those references surfaces $\Sigma$ are simply chosen to be $t=$ const. On any such spacelike hypersurface the integral necessarily diverges.

Instead, we can compute integrals only over the inner region and calculate the rate at which various charges change, that is, 
\be
Q(t) = \int_{\rom{inner}} J^\nu dS_v \label{charge_inner}.
\ee
The change of these charges should be  equal to outward flux crossing the neck region i.e.,
\be
\frac{dQ(t)}{dt} + \int_\rom{neck} J^r dS =0.
\ee

This is the computation we set-up in this section. We find that the inner region charges  decay monotonically in time. We therefore have the picture that the decay process corresponds to leakage of excitation from the inner region (AdS throat) to asymptotic infinity.   For simplicity we restrict our attention to $t=$ const hypersurfaces. Recall that in terms of the dimensional radial variable $x$, cf.~\eqref{eq:x-def}, the inner region is defined as $ x \ll \frac{1}{\epsilon^2}$. The neck is at $ x \sim \frac{1}{\epsilon^2}$.

Let us introduce a more convenient set of coordinates for computing integrals in the inner region. These coordinates also manifest the AdS nature of the inner region,
\be
x=\rho^{2}, \qquad \tau=\frac{t}{R}, \qquad \varphi=\frac{y}{R}. \label{coords_inner}
\ee
 In the large $R$ limit the inner region metric in these coordinates can be written  as \cite{Giusto:2012yz}\footnote{In this section we only focus on the six-dimensional part of the metric.},
\bea
ds^{2}_{6}&=&\sqrt{Q_{1}Q_{5}}\left[-\left(\rho^{2} + \frac{1}{k^{2}} \right) d\tau^{2} +  \left(\rho^{2} + \frac{1}{k^{2}} \right)^{-1}d\rho^{2}  + \rho^{2}d\varphi^{2} + d\theta^{2} \right. \nonumber\\
&+& \left. \sin^{2}\theta\left(d\phi\,+\,\frac{n}{k}\,d\varphi\,-\,\frac{n+1}{k}d\tau\right)^{2}\,+\,
\cos^{2}\theta\left(d\psi\,-\,\frac{n+1}{k}\,d\varphi\,+\,\frac{n}{k}d\tau\right)^{2}
\right]. \label{metric_inner}
\eea

 \subsection{Angular momenta of the perturbation in the inner region}
Angular momenta for the scalar perturbation associated to the $\psi$ and $\phi$ 
 are respectively,
\bea
L_\psi &=& \int T_\psi{}^\nu dS_\nu, \qquad\qquad 
L_\phi ~=~ \int T_\psi{}^\nu dS_\nu. \label{angular_momentum}
\eea
Substituting the separation ansatz \eqref{scalar_ansatz} in \eqref{stress_tensor}, we find the following expressions,
\begin{eqnarray}
 L_\psi &=&  2 m_\psi \int \sqrt{-g} dr d\mathcal{A} \left( -g^{tt} \omega_R + g^{t\psi} m_\psi + g^{t\phi} m_\phi + g^{ty} \lambda \right) \Phi \Phi^*,
\label{L_psi_simp}
\\
 L_\phi &=&  2 m_\phi \int \sqrt{-g} dr d\mathcal{A} \left( - g^{tt} \omega_R + g^{t\psi} m_\psi + g^{t\phi} m_\phi  + g^{ty} \lambda \right) \Phi \Phi^*,
\end{eqnarray}
where $d \cA = d \theta d\psi d\phi dy$. We note that the integrals in $L_\psi$ and $L_\phi$ are the same. It turns out that the integrals involved for the energy and the linear momentum are also the same. We focus on $L_\psi$; the discussion for other quantities is analogous. 

It is most convenient to compute this integral in the coordinates introduced in metric \eqref{coords_inner}. Let us recall that the scalar wavefunction in the inner region is given by, cf.~\eqref{scalar_ansatz}and~\eqref{hinner},
\be
\label{Phi_inner}
\Phi_\rom{in}=\rho^{k|\zeta|}\, \left(\rho^{2} + \frac{1}{k^2}\right)^\frac{{k\xi}}{2}\Theta(\theta)\,\rom{exp}{\left(-i\omega t + i\lambda y + im_{\psi}\psi + im_{\phi}\phi\right)} \nn
\left[{}_2F_1(a,b; c;- k^2\rho^{2})\right],
\ee
where the parameters in the arguments of the hypergeometric function ${}_2F_1$ are defined in eq.~\eqref{def_abc}. For the quasinormal modes, in the decoupling limit to leading order in $\epsilon=\frac{(Q_{1}Q_{5})^{\frac{1}{4}}}{R}$, we have that
\bea
\label{eqn26}\nu \simeq l+1\, \qquad \mbox{and} \qquad 1+\nu+ k|\zeta|-k\xi \simeq -2N.
\eea
As a result, 
\bea
(\Phi\Phi^{*})_\rom{in}&=&\rho^{2k|\zeta|}\left(\rho^{2} + \frac{1}{k^2}\right)^{k\xi}|\Theta(\theta)|^{2}\,e^{2\omega_It}\nonumber \\ 
& & \qquad \qquad \qquad \times  \left({}_2F_{1}(1+N+k|\zeta|,l+2+N+k|\zeta|;1+k|\zeta|;-k^{2}\rho^{2})\right)^{2}. \label{norm}
\eea
Using the inner region metric \eqref{metric_inner} we get
\bea
(L_{\psi})_\rom{in}\simeq \frac{2m_{\psi}Q_{1}Q_{5}}{R} \left(\omega_{R}R +\frac{n}{k}m_{\psi}-\frac{n+1}{k}m_{\phi}\right)\int_{0}^{\frac{1}{\epsilon}}\rho\,d\rho\int d\mathcal{A}\cos\theta\sin\theta\left(\rho^{2}+\frac{1}{k^{2}}\right)^{-1}(\Phi\Phi^{*})_\rom{in}.\nonumber\\
\eea
For small $\rho $, $\Phi_\rom{in} \sim \rho^{\,k|\zeta|}$. For large $\rho$, using \eqref{eqn26} and \eqref{inner_expansion},  we have $\Phi_\rom{in} \sim \rho^{-l-2}$, i.e.,  at this order in the approximation the wavefunction vanishes at large $\rho$.   As a result, in the large $\rho$ limit the integrand falls off as $\rho^{-2 l - 5}$. Therefore, to leading order in $\epsilon$ we can replace the upper limit of integration to infinity,
\bea
(L_{\psi})_\rom{in} \simeq \frac{2m_{\psi}Q_{1}Q_{5}}{R} \left(\omega_{R}R +\frac{n}{k}m_{\psi}-\frac{n+1}{k}m_{\phi}\right)\int_{0}^{\infty}\rho\,d\rho\int d\mathcal{A}\cos\theta\sin\theta\left(\rho^{2}+\frac{1}{k^{2}}\right)^{-1}(\Phi\Phi^{*})_\rom{in}.\nonumber\\
\eea
We now use the Euler transformation identity of the hypergeometric functions, equation 9.131.1 of~\cite{gradshteyn2007},
\bea
{}_2F_{1}(\alpha,\beta;\gamma;z)=(1-z)^{\gamma-\alpha-\beta}{}_2F_{1}(\gamma-\alpha,\gamma-\beta;\gamma;z),
\eea
to get
\bea
& & {}_2F_{1}(1+N+k|\zeta|,l+2+N+k|\zeta|;1+k|\zeta|;-k^{2}\rho^{2}) = (1+k^2\rho^{2})^{-2N-2-l-k|\zeta|} \nonumber \\
& & \qquad \qquad\qquad\qquad\qquad\qquad\qquad \qquad \qquad \times {}_2F_{1}(-N,-N-l-1;1+k|\zeta|;-k^{2}\rho^{2}).
\eea
Substituting this relation in the norm (\ref{norm})  we get for the inner region angular momentum,
\bea
(L_{\psi})_\rom{in}& \simeq & 4\pi C m_{\psi}Q_{1}Q_{5}\,e^{2\omega_{I}t}\left(\omega_{R}R +\frac{n}{k}m_{\psi}-\frac{n+1}{k}m_{\phi}\right) k^{-8N-8-4l-4k|\zeta|}  \nonumber\\
&&\int_{0}^{\infty} d\rho\,\rho^{2k{|\zeta|}+1}\left(\rho^{2}+\frac{1}{k^{2}}\right)^{-2N - l - k |\zeta| - 3}
\left({}_2F_{1}(-N,-N-l-1; 1+k|\zeta|; -k^{2}\rho^{2})\right)^{2},\nonumber\\
\eea
where $C = \int_{\rom{S}^3} d\theta d \phi d \psi \sin \theta \cos \theta |\Theta(\theta)|^2.$ Substituting $\tilde \rho = k \rho$, we get
\bea
(L_{\psi})_\rom{in}& \simeq & 4\pi C m_{\psi}Q_{1}Q_{5}\,e^{2\omega_{I}t}\left(\omega_{R}R +\frac{n}{k}m_{\psi}-\frac{n+1}{k}m_{\phi}\right) k^{-4N-4-2l-4k|\zeta|}  \nonumber\\
&&\int_{0}^{\infty} d\tilde \rho\,\tilde \rho^{2k{|\zeta|}+1}\left(\tilde \rho^2+1\right)^{-2N - l - k |\zeta| - 3}
\left({}_2F_{1}(-N,-N-l-1; 1+k|\zeta|; -\tilde \rho^{2})\right)^{2}.\nonumber\\
\eea
This integral can be evaluated using the hypergeometric function identity mentioned in appendix \ref{app:compendium}, equation \eqref{hypergeometric_identity}. Using that we get
\bea
(L_{\psi})_\rom{in}& \simeq & 4\pi C m_{\psi}Q_{1}Q_{5}\,e^{2\omega_{I}t}
\left(\omega_{R}R +\frac{n}{k}m_{\psi}-\frac{(n+1)}{k}m_{\phi}\right)k^{-4N-4-2 l- 4k|\zeta|}
\nonumber\\
&&\frac{\Gamma (1+k|\zeta|)^{2}\,\Gamma(N+1)\,\Gamma(N+l+2)}{2(2N+k|\zeta|+l+2)\Gamma (N+k|\zeta|+l+2)\,\Gamma(N+k|\zeta|+1)}.
\eea 
The first term in the denominator is simply $k\xi$, cf.~\eqref{eqn26}. Moreover, using equation
\eqref{xi_form} for $\xi$,
\bea
\xi \simeq   \omega_{R}R+\frac{n}{k}m_{\psi} - \frac{(n+1)}{k}m_{\phi},
\eea
 we see that the first term of the denominator cancels the term in the parenthesis. Thus, finally we get 
\bea
(L_{\psi})_\rom{in} \simeq 2\pi C\,Q_1Q_5\,m_{\psi}\,e^{2\omega_{I}t}
k^{-4N-5-2 l- 4k|\zeta|}
\,\frac{\Gamma (1+k|\zeta|)^{2}\,\Gamma(N+1)\,\Gamma(N+l+2)}{\Gamma (N+k|\zeta|+l+2)\,\Gamma(N+k|\zeta|+1)},
\eea
and as a result, 
\be
\frac{d}{dt} (L_{\psi})_\rom{in} =  2 \omega_I (L_{\psi})_\rom{in}.
\ee
Since $\omega_I$ is negative, the signs of  $(L_{\psi})_\rom{in}$ and $\frac{d}{dt} (L_{\psi})_\rom{in} $ are opposite.

The aim now is to recover this last expression from the flux of angular momentum across the neck region. 

\subsection{Flux of angular momenta of the perturbation across the neck}
\label{sec:flux}
To compute the outwards flux of angular momentum across the neck region, we need to compute the following integral, 
\be
F = \int_\rom{neck} J^r dS
\ee
where 
\be
dS = \sqrt{-g} d \theta d \phi d\psi dy = \sqrt{-g} d\cA = \sqrt{(Q_1 + f)(Q_5 + f)} \ r \sin \theta \cos\theta d\cA.
\ee
The radial component of angular momentum current is
\bea
J^r~~\equiv~~ T_\psi{}^{r} &=& \partial_\psi \Phi \partial^r  \Phi^* + \partial_\psi \Phi^* \partial^r  \Phi  \\ 
&=& i m_\psi (\Phi \partial^r  \Phi^* - \Phi^* \partial^r  \Phi) \\
&=& i m_\psi g^{rr} (\Phi \partial_r  \Phi^* - \Phi^* \partial_r  \Phi) \\
&=&  i m_\psi g^{rr} \left( \frac{\Phi\Phi^*}{H H^*} \right) \left(H(r) \partial_r H^*(r) - H^*(r) \partial_r H(r)\right).
\eea
Computing the expression in the second parenthesis for the neck region wavefunction \eqref{outer_expansion} for the quasinormal modes, we find
\bea
H(r) \partial_r H^*(r) - H^*(r) \partial_r H(r) &=& \frac{2 \nu |C_{2}|^{2}  a^2}{r^3\Gamma(1+\nu) \Gamma(1-\nu)} |\kappa|^{-2\nu} (\kappa^{2\nu} e^{-i \pi \nu} - (\kappa^*)^{2\nu} e^{i \pi \nu}).
\eea
To leading order in $\epsilon$, this expression becomes 
\bea
H(r) \partial_r H^*(r) - H^*(r) \partial_r H(r) \simeq - i \frac{4  |C_{2}|^{2} \sin^{2}(\pi\nu)  a^2}{\pi r^3}.
\eea
Therefore, we have
\be
J^r = 4 m_\psi a^2  \frac{g^{rr}}{\pi r^3}  e^{2 \omega_I t} |\Theta(\theta)|^2 |C_{2}|^{2} \sin^{2}(\pi\nu).
\ee
The inverse metric is written in appendix \ref{app:compendium}, equation \eqref{inverse_metric}. We have
\be
g^{rr} = \frac{r^2 + a^2 \gamma^2 \eta}{\sqrt{(Q_1 + f)(Q_5 +f)}}.
\ee
Using this expression, in the decoupling limit in the neck region we get, 
\be
J^r dS = 4 m_\psi a^2  \pi^{-1}  e^{2 \omega_I t} |\Theta(\theta)|^2 |C_{2}|^{2} \sin^{2}(\pi\nu) \sin \theta \cos \theta d\cA.
\ee
as a result flux is equal to, 
\be
F = 8  m_\psi R C   e^{2 \omega_I t} a^2 |C_{2}|^{2} \sin^{2}(\pi\nu). \label{flux_inter}
\ee

In order to have a final expression for this flux we need the  normalisation of the wavefunction $|C_{2}|^{2}$.

\subsection{Normalisation of the neck region wave function}
\label{sec:normalisation}

 On the one hand the outer region wave function in the neck region, $\kappa\sqrt{x} \ll 1 $,  reads as equation~\eqref{outer_expansion}. The no incoming waves boundary conditions at infinity relates $C_1$ and $C_2$ as \eqref{outgoing}. On the other hand,  the inner region wavefunction  in the neck region reads as equation~\eqref{inner_expansion}. Comparing the coefficients of $x^{\frac{\nu -1}{2}}$ of the above two expressions and using the relation \eqref{outgoing}, we get
\bea
\label{eq:norm_C_2}
C_2 = -  e^{i\nu\pi} \frac{k^{-1+\nu-k|\zeta|- k\xi}\Gamma(1+  k|\zeta|)\Gamma(\nu) \Gamma(1+\nu)}{\Gamma\left(\frac{1}{2}(1+  \nu +  k|\zeta| +  k\xi)\right)\Gamma\left(\frac{1}{2}(1+ \nu +  k|\zeta| -k\xi)\right)}\left(\frac{\kappa}{2}\right)^{-\nu}.
\eea

We are interested in modes where the second Gamma functions in the denominator of the above expression comes close to developing a pole, i.e., 
\bea
\label{eq:gamma_pole}
\Gamma\left(\frac{1}{2}(1+  \nu +  k|\zeta| -  k\xi)\right)=\Gamma(-N-\delta N)=\frac{(-1)^{N+1}}{N!}\frac{1}{\delta N}.
\eea
Here $\delta N$ is a small deviation from the integer $N$ that controls the divergence of the Gamma function.  We assume that $\delta N \ll \epsilon$ so that it can be ignored in the arguments of all the other Gamma functions. This is a consistent assumption~\cite{Chakrabarty:2015foa}.  An expression for $\delta N$ can be obtain from the matching condition \eqref{matchingnew2}.  Inserting such an expression for $\delta N$ in equation \eqref{eq:norm_C_2} we obtain the normalisation constant. We need to find $|C_{2}|^2$. A simple manipulation using these equations give,
\bea
\label{eqn42}
|C_{2}|^{2}\sin^{2}(\pi\nu)=\frac{\pi}{2}k^{-4 N - 5 - 4 k |\zeta| - 2l }(- \omega_I R)\frac{\Gamma(1+  k|\zeta|)^{2}\Gamma(N+1)\Gamma(N+\nu +1)}{\Gamma(N+\nu+1+k|\zeta|)\Gamma(N +1+k|\zeta|)},
\eea 
where $\omega_I$ is given in equation~\eqref{im_A_modes}. Substituting 
\eqref{eqn42} in \eqref{flux_inter} we indeed find that,
\be
\frac{d}{dt} (L_{\psi})_\rom{in} =  2 \omega_I (L_{\psi})_\rom{in} = - F.
\ee

To summarise: we have shown that the rate of change of angular momenta of the scalar perturbation  in the inner region can be matched with the fluxes of these quantities across the neck region. The angular momenta charges monotonically decay in the inner region with  decay exponent $\omega_I$. Therefore, we have picture that scalar excitation stuck in the AdS throat slowly leaks to asymptotic infinity.

 Very similar considerations apply to the energy and linear momentum of the scalar perturbation. The rate of change of energy and linear momentum in the inner region can be matched with the fluxes of these quantities across the neck region.

\section{Conclusions and discussion}
\label{sec:conclusions}
In this paper, we presented a detailed study of the quasinormal modes for the supersymmetric three-charge geometries of Giusto, Mathur, and Saxena (GMS)~\cite{gms1,gms2, Giusto:2004kj}. The key result of our paper is to reproduce the full spectrum in the decoupling limit from a D1-D5 orbifold CFT analysis. On the gravity side, our analysis differs slightly from the analysis of Eperon, Reall, and Santos (ERS)~\cite{Eperon:2016cdd}.  We work exclusively in the near decoupling limit. ERS work in the geometrical optics limit. In the regime of overlap, we reproduce their expressions for the real and imaginary parts of the quasinormal mode frequencies. On the CFT side, fortunately, we did not have to do much work. Correct interpretation of some of the earlier results in the literature \cite{Avery:2009tu, Avery:2009xr} allowed us to match the spectrum completely, including the orbifolding parameter. In section \ref{sec:wave_func} we studied properties of the scalar wavefunction and made precise the picture that the quasinormal modes in this set up represent slow leakage of excitation from AdS throat to asymptotic infinity.

There is a  technical observation we made in section \ref{sec:QNMs} that we wish to emphasise once again here. We observed that in 6d the GMS microstates have  a genuine  ergoregion but have no associated ergoregion instability.  This is somewhat surprising in both ways: the fact that a supersymmetric geometry has a genuine ergoregion and despite that it has no associated ergoregion instability. We are not aware of any other system where this happens. In contrast, let us mention that in the astrophysical relativity literature it is well known that all rotating stars suffer from the so-called Chandrasekhar-Friedman-Schutz (CFS) class of instability, see e.g.~\cite{Paschalidis:2016vmz}.  Ergoregion instability is such an instability. Such instabilities are thought to be generic and are in fact considered a significant hinderance in constructing truly stable rotating stars. In the present situation supersymmetry (and the Kaluza-Klein circle) does all that automatically for us, somehow.

In a recent series of papers \cite{Bianchi:2017sds, Bianchi:2018kzy, Bianchi:2019lmi}\footnote{We thank Guillaume Bossard for discussions on these papers.}, Bianchi et al  have studied geodesic motion  and properties of scalar scattering from fuzzballs closely related to the one studied above. They have computed impact parameters at which the massless particle moving on geodesics is captured by the fuzzball. They  show that the fuzzball geometries captures massless particles for a particular choice of impact parameter. This feature is once again different from properties of black holes. Black holes capture all geodesics impinging on them with impact parameter below a certain value. They also propose a qualitative picture of how the blackness property of fuzzball arises. It will be interesting to relate ours to  their study.  
 
The retarded bulk-to-bulk scalar Green's function on extremal BTZ black hole captures features related to Aretakis instability and associated power law decay~\cite{Ravishankar}.  A detailed study of the  bulk-to-bulk scalar Green's functions in asymptotically AdS microstate geometries, like the one presented in  \cite{Bena:2019azk} for the boundary-to-boundary scalar Green's function,  will be a good start in understanding the power law decay behaviour and any signature of Aretakis behavior in the fuzzball paradigm.

We end with reiterating~\cite{Bena:2013dka, Bena:2018mpb} the point that in the fuzzball paradigm one expects most of the microstate structure  to be intrinsically stringy. Our discussion of the ERS modes brings this point to the fore: we saw that within the scheme we worked with,  namely scalar excitation on the GMS microstate geometry, the dual CFT picture suggests transition to another state. Unfortunately, even for this rather simple set-up we do not know the bulk description of the new state beyond the linear approximation. Perhaps such a description is not possible within supergravity. Taking it one step further, the CFT picture  suggests  that the end state of the ERS instability is a transition to other (possibly stringy) mircostate structure. In our analysis on the CFT side, we do not see any feature related to formation of a tiny black hole as suggested by ERS; see also comments in~\cite{Marolf:2016nwu, Bena:2018mpb}. At this stage we do not know how to make this point more precise in the bulk description. On the CFT side perhaps more can be done, e.g., a model computation of moving from one state to another in a scalar scattering was done in~\cite{Lunin:2012gz}.

\subsection*{Acknowledgements}
We thank Guillaume Bossard, Sumanta Chakraborty, Oscar Dias, Suvrat Raju, Jorge Santos, David Turton, and especially  Iosif Bena and Samir Mathur for discussions. 
BC thanks CMI Chennai, TIFR Mumbai, IISER Pune, HRI Prayagraj, and IIT Gandhinagar for hospitality towards the final stages of this work. BC acknowledges the Infosys Program at ICTS for providing travel support to various conferences.  AV thanks AEI Potsdam for hospitality towards the final stages of this work. The work of AV and DG was supported in part by the Max Planck Partner Group ``Quantum Black Holes'' between CMI Chennai and AEI Potsdam. 

\appendix

\section{A compendium of formulae}
\label{app:compendium}
\subsection*{Inverse metric}
The inverse of the metric (\ref{extremalmetric}) is
\bea
&&g^{tt}=-{1\over h f}\Bigl(f+ Q_1+Q_5+Q_p+{Q_1 Q_5 + Q_1 Q_p + Q_5
Q_p\over
r^2+(\gamma_1+\gamma_2)^2\eta}\Bigr)\nonumber\\
&&g^{yy}={1\over h f}\Bigl(f+ Q_1+Q_5-Q_p+{Q_1 Q_5\,\eta\over r^2}-
{Q_p^2\,\eta\over r^2+(\gamma_1+\gamma_2)^2\eta}{(Q_1+Q_5)^2\over Q_1
Q_5}
\Bigr)\nonumber\\
&&g^{ty}=-{Q_p\over h f}\Bigl(1+{Q_1+Q_5\over
r^2+(\gamma_1+\gamma_2)^2\eta }
\Bigr)\nonumber\\
&&g^{\psi\psi}={1\over h f}\Bigl({1\over \cos^2\theta}+{\gamma_2^2\,
\eta\over r^2}-{\gamma_1^2\,\eta\over
r^2+(\gamma_1+\gamma_2)^2\eta}\Bigr)
\nonumber\\
&&g^{\phi\phi}={1\over h f}\Bigl({1\over \sin^2\theta}+{\gamma_1^2\,
\eta\over r^2}-{\gamma_2^2\,\eta\over
r^2+(\gamma_1+\gamma_2)^2\eta}\Bigr)
\nonumber\\
&&g^{\psi\phi}=-{Q_p\,\eta\over h f}\Bigl({1\over
r^2}-{1\over r^2+(\gamma_1+\gamma_2)^2\eta}\Bigr)\nonumber\\
&&g^{t\psi}=-{\sqrt{Q_1 Q_5}\over h f}{\gamma_1\over
r^2+(\gamma_1+\gamma_2)^2\eta}\nonumber\\
&&g^{t\phi}=-{\sqrt{Q_1 Q_5}\over h f}{\gamma_2\over
r^2+(\gamma_1+\gamma_2)^2\eta}\nonumber\\
&&g^{y\psi}={\sqrt{Q_1 Q_5} \gamma_2\,\eta\over h f}\Bigl({1\over r^2}+
{\gamma_1^2\over
r^2+(\gamma_1+\gamma_2)^2\eta}{Q_1+Q_5\over Q_1 Q_5}\Bigr)\nonumber\\
&&g^{y\phi}={\sqrt{Q_1 Q_5} \gamma_1\,\eta\over h f}\Bigl({1\over r^2}+
{\gamma_2^2\over
r^2+(\gamma_1+\gamma_2)^2\eta}{Q_1+Q_5\over Q_1 Q_5}\Bigr)\nonumber\\
&&g^{rr}={r^2 + (\gamma_1+\gamma_2)^2\eta\over h f}\,,\quad
g^{\theta\theta}={1\over h f}\,,\quad
g^{x_i x_j}= \sqrt{H_5\over H_1}\,\delta^{ij}. \label{inverse_metric}
\eea

\subsection*{A hypergeometric function identity}
For positive $\alpha $ and for arbitrary positive integers $N$ and $l$,
\bea
\label{hypergeometric_identity} 
\int_{0}^{\infty}d\rho\rho^{2\alpha+1}(1+\rho^{2})^{-2N-l-3-\alpha}
\left({}_2F_{1}(-N,-N-l-1,1+\alpha,-\rho^{2})\right)^{2}\nonumber\\
=\frac{1}{2(2N+\alpha + l +2)}\frac{\Gamma (1+\alpha)^{2}\,\Gamma(N+1)\,\Gamma(N+l+2)}{\Gamma (N+\alpha +l+2)\,\Gamma(N+\alpha+1)}.
\eea
A proof of the above identity can be found in appendix B.2 of \cite{Chakrabarty:2015foa}.

 \end{document}